\documentclass[prd,aps,twocolumn,showpacs,preprintnumbers,
superscriptaddress,floatfix,amssymb,eqsecnum]{revtex4} 
\usepackage{epsfig}
\usepackage{graphicx}
\usepackage{placeins}
\newcommand{\be}{\begin{equation}}
\newcommand{\ee}{\end{equation}}
\newcommand{\QchiPT}{Q$\chi$PT}
\newcommand{\chiPT}{$\chi$PT}

\begin{document}

\preprint{
\vbox{
\hbox{ADP-06-03/T634}
\hbox{Edinburgh 2006/09}
}}

\title{Precision Electromagnetic Structure of Octet Baryons in the
  Chiral Regime}

\author{S.~Boinepalli}
\affiliation{Special Research Centre for the Subatomic Structure of
             Matter and\\ Department of Physics, 
             University of Adelaide 5005, Australia}
\author{D.~B.~Leinweber}
\affiliation{Special Research Centre for the Subatomic Structure of
             Matter and\\ Department of Physics, 
             University of Adelaide 5005, Australia}
\author{A.~G.~Williams}
\affiliation{Special Research Centre for the Subatomic Structure of
             Matter and\\ Department of Physics, 
             University of Adelaide 5005, Australia}
\author{J.~M.~Zanotti}
\affiliation{Special Research Centre for the Subatomic Structure of
             Matter and\\ Department of Physics, 
             University of Adelaide 5005, Australia}
\affiliation{School of Physics, University of Edinburgh,
             Edinburgh EH9 3JZ, UK}
\author{J.~B.~Zhang}
\affiliation{Special Research Centre for the Subatomic Structure of
             Matter and\\ Department of Physics, 
             University of Adelaide 5005, Australia}
\affiliation{Zhejiang Institute of Modern Physics and Department of
             Physics, Zhejiang University, Hangzhou 310027, People's
	     Republic of China} 
\date{27 April 2006}

\begin{abstract}
  The electromagnetic properties of the baryon octet are calculated in
  quenched QCD on a $20^3 \times 40$ lattice with a lattice spacing of
  0.128~fm using the fat-link irrelevant clover (FLIC) fermion action.
  FLIC fermions enable simulations to be performed efficiently at
  quark masses as low as 300~MeV.
  By combining FLIC fermions with an improved-conserved vector
  current, we ensure that discretisation errors occur only at ${\cal
    O}(a^2)$ while maintaining current conservation.
  Magnetic moments and electric and magnetic radii are extracted from
  the electric and magnetic form factors for each individual quark
  sector.  From these, the corresponding baryon properties are
  constructed.
  Our results are compared extensively with the predictions of
  quenched chiral perturbation theory.
  We detect substantial curvature and environment sensitivity of the
  quark contributions to electric charge radii and magnetic moments in
  the low quark mass region.
  Furthermore, our quenched QCD simulation results are in accord with
  the leading non-analytic behaviour of quenched chiral perturbation
  theory, suggesting that the sum of higher-order terms makes only
  a small contribution to chiral curvature.
\end{abstract}

\pacs{12.39.Fe, 12.38.Gc, 13.40.Em, 14.20.Dh, 14.20.Jn}

\maketitle


\section{INTRODUCTION}
\label{sec:intro}

The study of the electromagnetic properties of baryons provides
valuable insight into the non-perturbative structure of QCD.  Baryon
charge radii and magnetic moments provide an excellent opportunity to
observe the chiral non-analytic behaviour of QCD.  Although the first
calculations of hadronic electromagnetic form factors appeared almost
20 years ago \cite{Wilcox:1985iy,Martinelli:1987bh,Draper:1989pi},
until recently the state-of-the-art calculations of the
electromagnetic properties of octet
\cite{Leinweber:1990dv,Wilcox:1991cq} and decuplet
\cite{Leinweber:1992hy} baryons and their electromagnetic transitions
\cite{Leinweber:1992pv} appeared almost 15 years ago.

However, over the last couple years there has been an increase in
activity in the area of octet baryon electromagnetic structure, mainly
by the Adelaide group
\cite{Zanotti:2004jy,Leinweber:2004tc,Leinweber:2005bz,Leinweber:2006ug} and the
QCDSF \cite{Gockeler:2003ay} and LHPC collaborations \cite{Edwards:2005kw}.
In this paper we improve upon our preliminary results reported in
Ref.~\cite{Zanotti:2004jy} and describe in detail the origin of the
lattice simulation results featured in
Refs.~\cite{Leinweber:2004tc,Leinweber:2005bz} and \cite{Leinweber:2006ug}
determining the strangeness magnetic moment and charge radius of the
nucleon respectively.

The extraction of baryon masses and electromagnetic form factors
proceeds through the calculation of Euclidean two and three-point
correlation functions, which are discussed at the hadronic level in
Section~\ref{subsec:CFHad}, and at the quark level in
Sections~\ref{subsec:CFQuark} and \ref{subsec:3ptQuark}.
Throughout this analysis we employ the lattice techniques introduced
in \cite{Leinweber:1990dv}.  We briefly outline the main aspects of these
techniques in section~\ref{sec:LatTech}.
The correlation functions directly proportional to the electromagnetic
form factors of interest are analysed in Sec.~\ref{sec:corrFun}.  
The results are presented and discussed in Section~\ref{sec:Results},
where an extensive comparison is made with the predictions of quenched
chiral perturbation theory (Q$\chi$PT)
\cite{Leinweber:2002qb,Savage:2001dy}.
A summary and discussion of future work is provided in
Section~\ref{sec:Summary}.


\section{THEORETICAL FORMALISM}
\label{sec:theory}

\subsection{Interpolating fields}
\label{subsec:Interpolators}

In this analysis we work with the standard established interpolating
fields commonly used in lattice QCD simulations.
The notation adopted is similar to that of \cite{Leinweber:1990dv}.
To access the proton we use the positive parity interpolating field
\begin{eqnarray}
\label{chi1p}
\chi^{p +}(x)
&=& \epsilon^{abc}
\left( u^{aT}(x)\ C \gamma_5\ d^b(x) \right) u^c(x)\ ,
\end{eqnarray}
where the fields $u$, $d$ are evaluated at Euclidean space-time point
$x$, $C$ is the charge conjugation matrix, $a,\ b$ and $c$ are colour
labels, and the superscript $T$ denotes the transpose.
In this paper we follow the notation of Sakurai.  The Dirac $\gamma$
matrices are Hermitian and satisfy $\{\gamma_\mu,\gamma_\nu\} =
2\delta_{\mu\nu}$, with $\sigma_{\mu\nu} =
(1/2i)[\gamma_\mu,\gamma_\nu]$.
This interpolating field transforms as a spinor under a parity
transformation.
That is, if the quark fields $q^a(x)\ (q=u,d, \cdots)$ transform as
\begin{equation}
{\cal P}\, q^a(x)\, {\cal P}^\dagger = +\gamma_0\, q^a(\tilde{x})\ ,
\end{equation}
where $\tilde{x} = (x_0 , -\vec{x})$, then
\begin{equation}
{\cal P}\, \chi^{p +}(x)\, {\cal P}^\dagger
 = +\gamma_0\, \chi^{p +}(\tilde{x})\ .
\end{equation}

The neutron interpolating field is obtained via the exchange
$u\leftrightarrow d$, and the strangeness $-2$, $\Xi$ interpolating
fields are obtained by replacing the doubly represented $u$ or $d$
quark fields in Eq.~(\ref{chi1p}) by $s$.  Similarly, the charged
strangeness $-1$, $\Sigma$ interpolating fields are obtained by
replacing the singly represented $u$ or $d$ quark fields in
Eq.~(\ref{chi1p}) by $s$.
For the $\Sigma^0$ hyperon one uses \cite{Leinweber:1990dv}
\begin{eqnarray}
\chi^{\Sigma^0}(x)
= {1\over\sqrt{2}} \epsilon^{abc}
\!\!\!&\Big\{&\!\!\! \left( u^{aT}(x)\ C \gamma_5\ s^b(x) \right) d^c(x)\nonumber \\
     \!\!\!&+&\!\!\! \left( d^{aT}(x)\ C \gamma_5\ s^b(x) \right) u^c(x)
\Big\}\, , 
\label{chi1S} 
\end{eqnarray}
Note that $\chi^{\Sigma^0}$ transforms as a triplet under SU(2)
isospin.  
An SU(2) isosinglet interpolating field for the $\Lambda$ can be
constructed by replacing $``+" \longrightarrow ``-"$ in
Eq.~(\ref{chi1S}).
For the SU(3) octet $\Lambda$ interpolating field, one has
\begin{widetext}
\begin{equation}
\chi^{\Lambda}(x)
= {1\over\sqrt{6}} \epsilon^{abc}
\Big\{ 2
   \left( u^{aT}(x)\ C \gamma_5\ d^b(x) \right) s^c(x)
+ \left( u^{aT}(x)\ C \gamma_5\ s^b(x) \right) d^c(x)\
- \left( d^{aT}(x)\ C \gamma_5\ s^b(x) \right) u^c(x)
\Big\}\, .
\label{chi1l8}
\end{equation}
We select this interpolating field for studying the $\Lambda$ in the
following. 

%

\subsection{Correlation functions at the hadronic level}
\label{subsec:CFHad}

The extraction of baryon masses and electromagnetic form factors
proceeds through the calculation of the ensemble average (denoted
$\bigm < \cdots \bigm >$) of two and three-point correlation
functions.
The two-point function is defined as
\begin {equation}
 \bigm < G^{BB} (t;\vec{p}, \Gamma) \bigm > = \sum_{\vec{x}} e^{-i
\vec{p} \cdot \vec{x} }\, \Gamma^{\beta \alpha} \, \bigm < \Omega \bigm | T
\left ( \chi^\alpha(x)\, \overline \chi^\beta(0) \right )
\bigm | \Omega \bigm >\ .
\label{twoptfn}
\end {equation}
Here $\Omega$ represents the QCD vacuum, $\Gamma$ is a $4 \times 4$
matrix in Dirac space and $\alpha, \, \beta$ are Dirac indices. 
At the hadronic level we insert a complete set of states $\bigm | B,
p, s \bigm >$ and define
\begin{equation}
 \bigm < \Omega \bigm | \chi (0) \bigm | B, p, s \bigm > \, =
   Z_B(p) \sqrt\frac{M}{E_p} \, u(p,s)\ ,
\label{omega}
\end{equation}
where $Z_B(p)$ represents the  coupling strength of
$\chi(0)$ to baryon $B$, and $E_p = \sqrt{\vec{p}^2 + M^2}$.  A
momentum dependence for $Z_B(p)$  is included for the case where a
smeared sink is employed.
For large Euclidean time 
\begin{equation}
 \bigm < G^{BB} (t;\vec{p}, \Gamma) \bigm > \, \simeq
 \frac{Z_B(p)\overline{Z}_B(p)}{2 E_p}
   e^{- E_p t} \, tr \left [ \Gamma (-i \gamma \cdot p + M) \right ] 
 \ .
\label{twoptfnsimp}
\end{equation}
Here $\overline{Z}_B(p)$ is the coupling strength of the source
$\overline\chi(0)$ to the baryon.  Again, the momentum dependence
allows for the use of smeared fermion sources in the creation of the
quark propagators and the differentiation between source and sink
allows for our use of smeared sources and point sinks in the
following.
Similarly the three-point correlation function for the electromagnetic
current, $j^\mu(x)$, is defined as
\begin{equation}
{\quad \bigm < G^{B j^\mu B} 
 (t_2, t_1; \vec{p'}, \vec{p}; \Gamma) \bigm >  = \sum_{\vec{x_2},
   \vec{x_1}}
   e^{-i \vec{p'} \cdot \vec{x_2} } e^{+ i \left (
   \vec{p'} - \vec{p} \right ) \cdot \vec{x_1} } 
   \Gamma^{\beta \alpha} \bigm < \Omega \bigm | T
   \left ( \chi^\alpha(x_2) j^\mu(x_1) \overline \chi^\beta(0) \right ) 
   \bigm | \Omega \bigm >\ .\quad }
\label{threeptfn}
\end{equation}
For large Euclidean time separations $t_2 - t_1 >\!> 1$ and $t_1 >\!>
1$, the three-point function at the hadronic level is dominated by the
contribution from the ground state
\begin{equation}
{\quad \bigm < G^{B j^\mu B} 
   (t_2, t_1;\vec{p'}, \vec{p}; \Gamma) \bigm > 
   = \sum_{s, s^\prime}
   e^{-E_{p'} (t_2-t_1)} e^{-E_p t_1} \Gamma^{\beta \alpha} \bigm <
   \Omega \bigm | \chi^\alpha \bigm |
   p', s^\prime \bigm > \bigm < p', s^\prime \bigm |
   j^\mu \bigm | p, s \bigm > \bigm < p, s \bigm | 
   \overline \chi^\beta  \bigm | \Omega \bigm >\ . \quad
 }
\label{threeptfnhad}
\end{equation}
The matrix element of the electromagnetic current has the general form
\begin{equation}
 \bigm < p', s^\prime \bigm | j^\mu \bigm | p, s \bigm > \, =
   \left ( M^2 \over E_p E_{p'} \right )^{1/2} 
   \overline u (p', s^\prime) 
   \left ( F_1(q^2) \gamma^\mu - 
           F_2(q^2) \sigma^{\mu \nu} {q^\nu \over 2 M} \right )
    u(p, s)\ ,
\label{emcurrent}
\end{equation}
where $q= p' - p$. 
To eliminate the time dependence of the three-point functions we
construct the following ratio,
\begin{equation}
   R(t_2,t_1; \vec{p'}, \vec{p}; \Gamma, \Gamma^\prime; \mu )
   = \left [\frac {
\bigm < G^{B j^\mu B} (t_2,t_1; \vec{p'}, \vec{p}; \Gamma) \bigm >
\bigm < G^{B j^\mu B} (t_2,t_1; -\vec{p}, -\vec{p'}; \Gamma) \bigm >}
{\bigm < G^{BB} (t_2; \vec{p'}; \Gamma^{'} ) \bigm >
\bigm < G^{BB} (t_2; -\vec{p}; \Gamma^{'} ) \bigm >} 
\right ]^{1/2}\ . 
\label{ratio}
 \end{equation}
For large time separations $t_2-t_1 >\!> 1$ and $t_1 >\!> 1$ this
ratio is constant in time and is proportional to the
electromagnetic form factors of interest.
We further define a reduced ratio
$\overline{R}(\vec{p^{'}},\vec{p};\Gamma,\Gamma^{'};\mu)$ as
\begin{equation}
\overline{R}(\vec{p^{'}},\vec{p};\Gamma,\Gamma^{'};\mu) =  \left
  [\frac{2E_p}{E_p+M} \right ]^{1/2} \left [\frac{2E_{p^{'}}}
{E_{p^{'}}+M} \right ]^{1/2}
R(t_2,t_1; \vec{p'}, \vec{p}; \Gamma, \Gamma^{'}; \mu )\ ,
\label{ratioreduced}
\end{equation}
\end{widetext}
from which the Sachs forms for the electromagnetic form factors
\begin{eqnarray}
 {\cal G}_E(q^2) & = &  F_1(q^2) - {\frac{q^2}{(2M)^2}} F_2(q^2)
\ ,\\
 {\cal G}_M(q^2) & = & F_1(q^2) + F_2(q^2)\ ,
\label{sachs}
\end{eqnarray}
may be extracted through an appropriate choice of $\Gamma$ and
$\Gamma^{'}$.
A straight forward calculation reveals
\begin{eqnarray}
{\cal G}_E(q^2) & = &   \overline R(\vec{q},\vec{0}; \Gamma_4,
     \Gamma_4, 4) \, , \quad 
\label{GEratio} \\
{| \epsilon_{ijk} q^i |} \, {\cal G}_M(q^2) & = & (E_q + M) \,
  \overline R(\vec{q},\vec{0}; \Gamma_j, \Gamma_4, k) \, , \quad
\label{GMratio} \\
{|q^k|}\, {\cal G}_E(q^2) & = &  (E_q + M)\, 
  \overline R(\vec{q},\vec{0}; \Gamma_4, \Gamma_4, k) \, , \quad
\label{GEratio2}
\end{eqnarray}
where
\begin{eqnarray}
 \Gamma_j & = & \frac{1}{2} \left( \begin{array}{lr} \sigma_j & 0  \\
 0 & 0 \end{array} \right)\, ,  \nonumber \\
 \Gamma_4 & = & \frac{1}{2} \left ( \begin{array}{lr}
                                           I & 0 \\
                                           0 & 0 
                                    \end{array}  \right )\, .
\label{gamma}
\end{eqnarray}

\subsection{Correlation functions at the quark level}
\label{subsec:CFQuark}

Here the two and three-point functions of Sec.~\ref{subsec:CFHad} are
calculated at the quark level by using the explicit forms of the
interpolating fields of Sec.~\ref{subsec:Interpolators} and
contracting out all possible pairs of quark field operators.  These
become quark propagators in the ensemble average.  For convenience, we
introduce the shorthand notation for the correlation functions ${\cal
  G}$ of quark propagators $S$
\begin{widetext}
\be
{\cal G}(S_{f_1},S_{f_2},S_{f_3}) \equiv
   \epsilon^{abc} \epsilon^{a'b'c'}
  \biggl\{ 
   S_{f_1}^{a a'}(x,0) \, {\rm tr} \left [ S_{f_2}^{b b' \, T}(x,0)
   S_{f_3}^{c c'}(x,0)
   \right ] + S_{f_1}^{a a'}(x,0) \, S_{f_2}^{b b'
     \, T}(x,0) \, S_{f_3}^{c c'}(x,0) \biggr \}\ ,
\label{F}
\ee
where $S^{a a'}_{f_{1-3}}(x,0)$ are the quark propagators in the
background link-field configuration $U$ corresponding to flavours
$f_{1-3}$.
This allows us to express the correlation functions in a compact form.
The associated correlation function for $\chi^{p+}$ can be written
as
\begin{equation}
G^{p+}(t,\vec p; \Gamma) = \left\langle \sum_{\vec x}
  e^{-i \vec p \cdot \vec x} {\rm tr}
  \left[ \Gamma \, \, {\cal G}
    \left( S_u, \, \widetilde C S_d {\widetilde C}^{-1}, \, S_u
    \right)
  \right] \right\rangle\ ,
\label{p11CF}
\end{equation}
where $\langle\cdots\rangle$ is the ensemble average over the link
fields, $\Gamma$ is the $\Gamma_{\pm}$ projection operator that
separates the positive and negative parity states, and $\widetilde C =
C\gamma_5$.
For ease of notation, we will drop the angled brackets,
$\langle\cdots\rangle$, and all the following correlation functions
will be understood to be ensemble averages.

Two-point correlation functions for other octet baryons composed of a
doubly-represented quark flavour and a singly-represented quark
flavour follow from Eq.~(\ref{p11CF}) with the appropriate
substitution of flavour subscripts.  The correlation function for the
neutral member $\Sigma^0$ is given by the average of correlation
functions for the charged states $\Sigma^+$ and $\Sigma^-$.  Finally
the correlation function for $\Lambda$ obtained from the
octet-interpolating field of Eq.~(\ref{chi1l8}) is
\begin{eqnarray}
G^{\Lambda^8}(t,\vec p; \Gamma)  =
  {1 \over 6} \sum_{\vec x} e^{-i \vec p \cdot \vec x}
   {\rm tr} \biggl [ \Gamma
\!\!  &\biggl \{& \!\!
     2 \,  {\cal G} \left ( S_s, \,\widetilde C S_u
   \widetilde C^{-1}, \, S_d \right )
   + 2 \, {\cal G} \left ( S_s, \, \widetilde C S_d
    \widetilde C^{-1}, \, S_u \right ) \nonumber\\
\!\! &+& \!\! 2 \, {\cal G} \left ( S_d, \, \widetilde C S_u
    \widetilde C^{-1}, \, S_s \right )
   + 2 \, {\cal G} \left ( S_u, \, \widetilde C S_d
    \widetilde C^{-1}, \, S_s \right ) \nonumber\\
\!\! &-& \!\! \phantom{2} \, {\cal G} \left ( S_d, \, \widetilde C S_s
    \widetilde C^{-1}, \, S_u \right )
   - \phantom{2} \, {\cal G} \left ( S_u, \, \widetilde C S_s
    \widetilde C^{-1}, \, S_d \right )
   \biggr \} \biggr ]\ .
\label{L118CF}
\end{eqnarray}
\end{widetext}

\subsection{Three-point functions at the quark level}
\label{subsec:3ptQuark}

\begin{figure}[tbp]
{\includegraphics[height=3.3cm,angle=90]{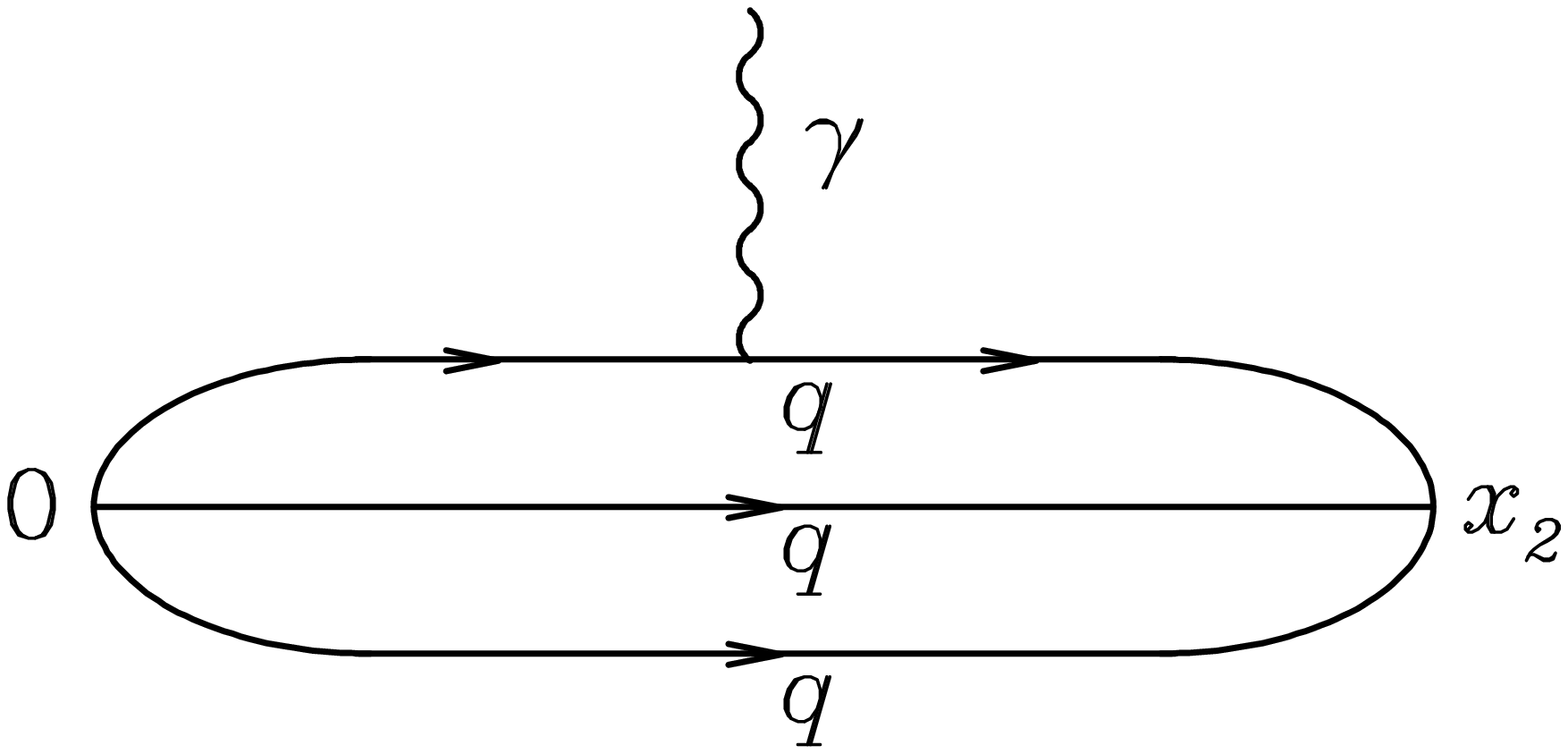} \hspace{0.8cm}
 \includegraphics[height=3.3cm,angle=90]{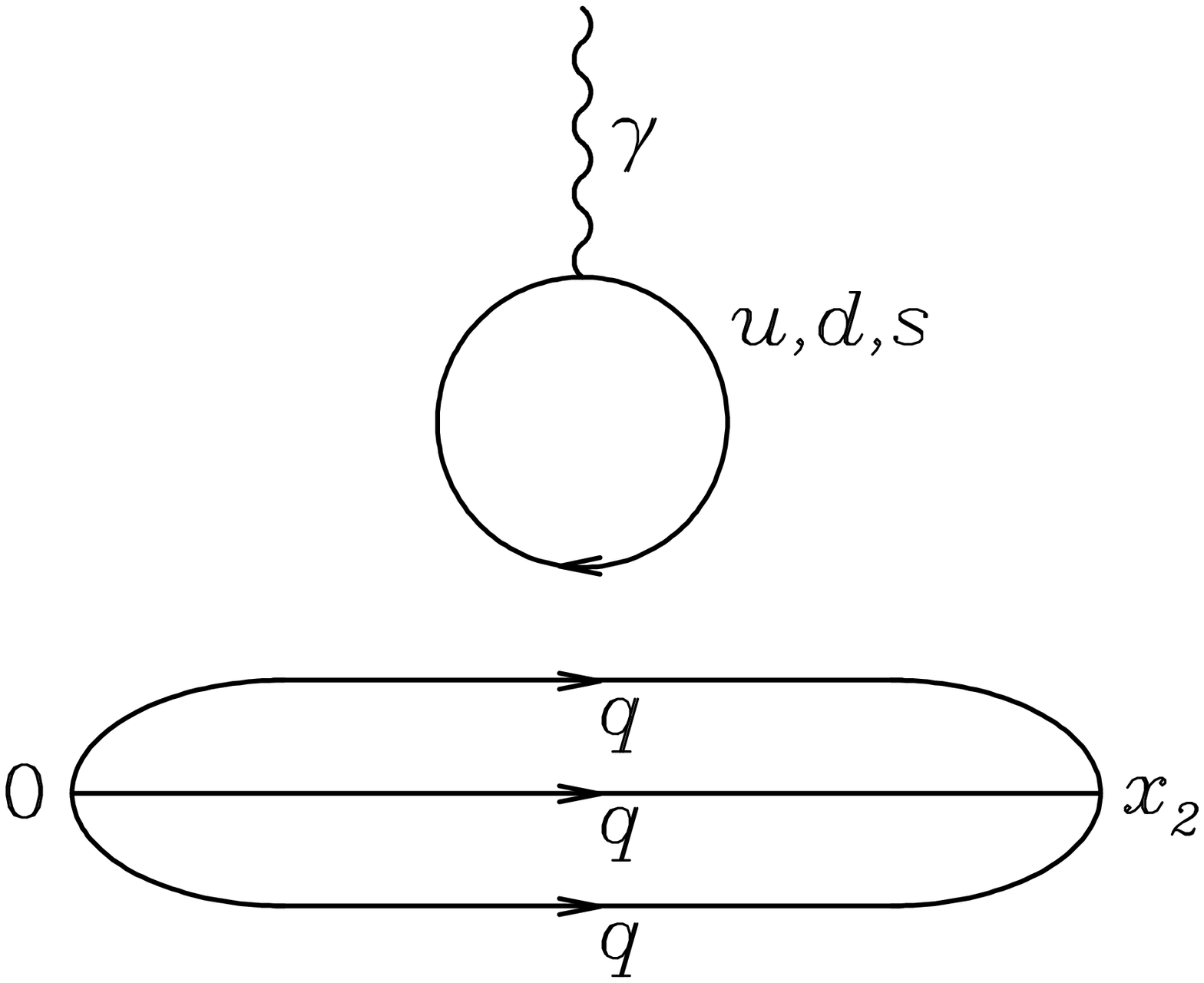}}
\caption{Diagrams illustrating the two topologically different
insertions of the current within the framework of lattice QCD.  
}
\label{topology}
\end{figure}

   In determining the three point function, one encounters two
topologically different ways of performing the current insertion.
Figure \ref{topology} displays skeleton diagrams for these two
insertions.  These diagrams may be dressed with an arbitrary number of
gluons (and additional sea-quark loops in full QCD).  Diagram (a)
illustrates the connected insertion of the current to one of the
quarks created via the baryon interpolating field.  This simple
skeleton diagram does indeed contain a sea-quark component, as upon
dressing the diagram with gluon exchange, quark-loop and $Z$-diagrams
flows become possible.  It is here that ``Pauli-blocking'' in the sea
contributions, central to obtaining violation of the Gottfried sum
rule, are taken into account.  Diagram (b) accounts for an alternative
quark-field contraction where the current first produces a
disconnected $q \, \overline q$ loop-pair which in turn interacts with
the valence quarks of the baryon via gluons.

Thus, the number of terms in the three-point function is four times
that in Eq.~(\ref{p11CF}).  The correlation function for proton matrix
elements obtained from the interpolator of Eq.~(\ref{chi1p}) is
\begin{widetext}
\begin{eqnarray}
\lefteqn{
T \left ( \chi^{p+}(x_2) \, j^\mu(x_1) \,
\overline \chi^{\, p+}(0) \right )  = 
} 
\qquad\qquad \nonumber \\
&& \;\;\:
{\cal G}\left( \widehat S_u(x_2,x_1,0), \, \widetilde C 
                        S_d(x_2,0) {\widetilde C}^{-1}, \, 
                        S_u(x_2,0) \right) +
{\cal G}\left(          S_u(x_2,0), \, \widetilde C 
                        S_d(x_2,0) {\widetilde C}^{-1}, \, 
               \widehat S_u(x_2,x_1,0) \right)
\nonumber \\
&& +
{\cal G}\left(          S_u(x_2,0), \, \widetilde C 
               \widehat S_d(x_2,x_1,0) {\widetilde C}^{-1}, \, 
                        S_u(x_2,0) \right) 
\nonumber \\
&& + \sum_{q = u,\, d,\, s} e_q\, \sum_i
tr \left [ S_q^{ii}(x_1,x_1) \, \gamma_\mu \right ] 
{\cal G}\left( S_u(x_2,0), \, \widetilde C 
               S_d(x_2,0) {\widetilde C}^{-1}, \, 
               S_u(x_2,0) \right) \, ,
\label{Prot3pt}
\end{eqnarray}
\end{widetext}
where
\begin{equation}
\widehat S_q^{aa'}(x_2,x_1,0) = e_q\, \sum_{i} S_q^{ai}(x_2,x_1) \,
\gamma_\mu \, S_q^{i a'}(x_1,0) \, ,
\label{ConnIns}
\end{equation}
denotes the connected insertion of the electromagnetic current to a
quark of charge $e_q$.

The first two terms of Eq.~(\ref{Prot3pt}) provide the connected insertion
contribution of the u-quark sector to the proton's electromagnetic
properties, whereas the third term provides the connected $d$-quark
contribution.  The latter term of Eq.~(\ref{Prot3pt}) accounts for the
``disconnected'' loop contribution depicted in Fig.~\ref{topology}b.
Here, the sum over the quarks running around the loop has been
restricted to the flavours relevant to the ground state baryon octet.
In the SU(3)-flavour limit the sum vanishes for the electromagnetic
current.  However, the heavier strange quark mass allows for a
non-trivial result.

The ``disconnected'' current insertion requires a numerical estimate
of $S_q^{ii}(x_1,x_1)$ for the lattice volume of diagonal spatial
indices.  As this requires numerous source vectors in the
fermion-matrix inversion, determination of this propagator is
numerically intensive \cite{Mathur:2000cf,Lewis:2002ix,Foley:2005ac}.
Indeed, an indirect method using experimental results, chiral
effective field theory and the lattice results from the connected
current insertion presented herein, provides the most precise
determinations of these quark loop contributions to the nucleon's
electromagnetic structure
\cite{Leinweber:2004tc,Leinweber:2005bz,Leinweber:2006ug} at present.
This approach should be viewed as complementary to an {\it ab
initio} determination via lattice QCD which awaits a next-generation
dynamical-fermion simulation of QCD \cite{Leinweber:2002qb}.

It is interesting to examine the structure of the connected insertion
contributions to the proton's structure.  Here, we see very different
roles played by $u$ and $d$ quarks in the correlation function, in
that only the $d$-quark appears in the second position of $\cal G$.
The absence of equivalence for $u$ and $d$ contributions allows the
connected quark sector to give rise to a non-trivial neutron charge
radius, a large neutron magnetic moment, or a violation of the
Gottfried sum rule.  As each term of Eq.~(\ref{Prot3pt}) can be
calculated individually, it is a simple task to isolate the quark
sector contributions to the baryon electromagnetic properties.

Another interesting point to emphasise, is that there is no simple
relationship between the properties of a particular quark flavour bound
in different baryons.  For example, the correlator for $\Sigma^+$ is
given by (\ref{Prot3pt}) with $d \to s$.  Hence, a $u$-quark
propagator in $\Sigma^+$ is multiplied by an $s$-quark propagator,
whereas in the proton the $u$-quark propagators are multiplied by a
$d$-quark propagator.  The different mass of the neighbouring quark
gives rise to an environment sensitivity in the $u$-quark
contributions to
observables\cite{Leinweber:1990dv,Leinweber:1991vc,Leinweber:1992hy,%
Leinweber:1992pv,Leinweber:1993nr,Leinweber:1999nf,Leinweber:2004tc}.
This point sharply contrasts the concept of an intrinsic quark
property which is independent of the quark's environment.  This
concept of an intrinsic quark property is a fundamental foundation of
many constituent based quark models and is not in accord with QCD.

%
\section{LATTICE TECHNIQUES}
\label{sec:LatTech}

\subsection{Gauge and quark actions}

The simulations are performed using the mean-field ${\cal
  O}(a^2)$-improved Luscher-Weisz \cite{Luscher:1984xn} plaquette plus
rectangle gauge action on a $20^3 \times 40$ lattice with periodic
boundary conditions.  The lattice spacing $a=0.128$ fm is determined
by the Sommer scale $r_0=0.50$ fm \cite{Sommer:1993ce}.  This large
volume lattice ensures a good density of low-lying momenta which are
key to giving rise to chiral non-analytic behaviour in the observables
simulated on the lattice
\cite{Leinweber:2004tc,Leinweber:2005bz,Leinweber:2006ug}.

We perform a high-statistics analysis using a large sample of 400
configurations for our lightest eight quark masses.  We also consider
a subset of 200 configurations for our three heaviest quark masses to
explore the approach to the heavy-quark regime.  A sub-ensemble bias
correction is applied multiplicatively to the heavy quark results, by
matching the central values of the 200 configuration sub-ensemble and
400 configuration ensemble averages at $\kappa = 0.12780$.
The error analysis is performed by a third-order, single-elimination
jackknife.

For the quark fields, we use the Fat-Link Irrelevant Clover fermion
action \cite{FATJAMES}
\be
S_{\rm SW}^{\rm FL}
= S_{\rm W}^{\rm FL} - \frac{i\, g\, C_{\rm SW}\, \kappa\,
  r}{2(u_{0}^{\rm FL})^4}\, 
             \bar{\psi}(x)\, \sigma_{\mu\nu}F_{\mu\nu}\, \psi(x)\, ,
\label{FLIC}
\ee
where $F_{\mu\nu}$ is an ${\cal O}(a^4)$-improved lattice definition
\cite{Bilson-Thompson:2002jk} constructed using fat links and
$u_{0}^{\rm FL}$ is the plaquette measure of the mean link calculated
with fat links.  The mean-field improved Fat-Link Irrelevant Wilson
action is
\begin{eqnarray}
S_{\rm W}^{\rm FL}
=  &\sum_x& \bar{\psi}(x)\psi(x) 
+ \kappa \sum_{x,\mu} \bar{\psi}(x)
    \bigg[ \gamma_{\mu}
      \bigg( \frac{U_{\mu}(x)}{u_0} \psi(x+\hat{\mu}) \nonumber \\
&-& \frac{U^{\dagger}_{\mu}(x-\hat{\mu})}{u_0} \psi(x-\hat{\mu})
      \bigg)
- r \bigg(
 \frac{U_{\mu}^{\rm FL}(x)}{u_0^{\rm  FL}} \psi(x+\hat{\mu})\nonumber\\
&+& \frac{U^{{\rm FL}\dagger}_{\mu}(x-\hat{\mu})}{u_0^{\rm FL}}
          \psi(x-\hat{\mu})
      \bigg)
    \bigg]\ .
\label{MFIW}
\end{eqnarray}
with $\kappa = 1/(2m+8r)$. We take the standard value $r=1$.  
Our notation uses the Pauli representation of the Dirac
$\gamma$-matrices \cite{SAKURAI}, where the $\gamma$-matrices are
hermitian and $\sigma_{\mu\nu} = [\gamma_{\mu},\ \gamma_{\nu}]/(2i)$.
Fat links are constructed by performing $n_{\rm APE}=6$ sweeps of APE
smearing, where in each sweep the weights given to the original link
and the six transverse staples are 0.3 and $(0.7/6)$ respectively.
The FLIC action is closely related to the mean-field improved clover
(MFIC) fermion action in that the latter is described by
Eqs.~(\ref{FLIC}) and (\ref{MFIW}) with all fat-links replaced by
untouched thin links and $F_{\mu\nu}$ defined by the $1 \times 1$-loop
clover definition.

For fat links, the mean link $u_0 \approx 1$, indicating that
perturbative renormalisations are small for smeared links and are
accurately accounted for by small mean-field improvement corrections.
As a result, mean-field improvement of the coefficients of the clover
and Wilson terms of the fermion action is sufficient to accurately
match these terms and eliminate ${\cal O}(a)$ errors from the fermion
action \cite{Zanotti:2004dr}.
An added advantage is that access to the light quark mass regime is
enabled by the improved chiral properties of the FLIC fermion
action \cite{FLIClqm}.

Time slices are labeled from 1 to 40, and a fixed boundary condition
at $t=40$ is used for the fermions.  An analysis of the pion
correlator indicates that the effects of this boundary condition are
negligible for $t \le 30$, and all of our correlation-function fits
are performed well within this regime.

Gauge-invariant Gaussian smearing \cite{Gusken:qx,Zanotti:2003fx} in
the spatial dimensions is applied at the source at $t=8$ to increase
the overlap of the interpolating operators with the ground state while
suppressing excited state contributions.

Table~\ref{tab:baryonmasses} provides the kappa values used in our
simulations, together with the calculated $\pi$ and octet baryon
masses.  While we refer to $m_\pi^2$ to infer the quark masses, we
note that the critical value where the pion mass vanishes is
$\kappa_{\rm cr} = 0.13135$.

We select $\kappa = 0.12885$ to represent the strange quark in this
simulation.  At this $\kappa$ the $s\bar s$ pseudoscalar mass is
0.697~GeV, which compares well with the experimental value of $2\, m_{\rm
K}^2 - m_\pi^2 = ( 0.693\ {\rm GeV} )^2$, motivated by leading order
chiral perturbation theory.

\begin{table*}[tbp]
\caption{Hadron masses in appropriate powers of GeV for various values
  of the hopping parameter, $\kappa$.  For reference, experimentally
  measured values are indicated at the end of the table.}
\label{tab:baryonmasses}
\begin{ruledtabular}
\begin{tabular}{cccccc}
\noalign{\smallskip}
$\kappa$  & $m_\pi^2$    & $N$          & $\Lambda$   & $\Sigma$     & $\Xi$  \\
\hline
\noalign{\smallskip}
$0.12630$ & $0.9972(55)$ & $1.829(8) $  & $1.728(10)$ & $1.700(9)$   & $1.612(11)$ \\
$0.12680$ & $0.8947(54)$ & $1.763(9) $  & $1.681(10)$ & $1.656(10)$  & $1.586(12)$ \\
$0.12730$ & $0.7931(53)$ & $1.695(9) $  & $1.632(11)$ & $1.566(11)$  & $1.558(12)$ \\
$0.12780$ & $0.6910(35)$ & $1.629(10)$  & $1.584(10)$ & $1.570(10)$  & $1.531(10)$ \\
$0.12830$ & $0.5925(33)$ & $1.554(10)$  & $1.530(10)$ & $1.521(10)$  & $1.502(10)$ \\
$0.12885$ & $0.4854(31)$ & $1.468(11)$  & $1.468(11)$ & $1.468(11)$  & $1.468(11)$  \\
$0.12940$ & $0.3795(31)$ & $1.383(11)$  & $1.406(11)$ & $1.417(11)$  & $1.435(11)$  \\
$0.12990$ & $0.2839(33)$ & $1.301(11)$  & $1.347(11)$ & $1.371(11)$  & $1.404(11)$ \\
$0.13025$ & $0.2153(35)$ & $1.243(12)$  & $1.303(12)$ & $1.341(12)$  & $1.382(11)$  \\
$0.13060$ & $0.1384(43)$ & $1.190(15)$  & $1.256(13)$ & $1.313(12)$  & $1.359(11)$ \\
$0.13080$ & $0.0939(44)$ & $1.159(21)$  & $1.226(16)$ & $1.296(14)$  & $1.346(11)$ \\
experiment & 0.0196 & $0.939$ & $1.116$ & $1.189$  & $1.315$ \\
\end{tabular}
\end{ruledtabular}
\end{table*}

\subsection{Improved conserved vector current}

For the construction of the ${\cal O}(a)$-improved conserved vector
current, we follow the technique proposed by Martinelli {\it et al.}
\cite{Martinelli:ny}.  The standard conserved vector current for
Wilson-type fermions is derived via the Noether procedure
\begin{eqnarray}
j_\mu^{\rm C} &\equiv& \frac{1}{4}\bigl[\overline{\psi}(x) (\gamma_\mu -
r)U_\mu(x) \psi(x+\hat{\mu}) \nonumber \\ 
&+& \overline{\psi}(x+\hat{\mu}) (\gamma_\mu + r)U_\mu^\dagger(x)
\psi(x) \nonumber \\
&+& (x\rightarrow x-\hat{\mu})\bigr] .
\label{conserved}
\end{eqnarray}
The ${\cal O}(a)$-improvement term is also derived from the fermion
action and is constructed in the form of a total four-divergence,
preserving charge conservation.  The ${\cal O}(a)$-improved conserved
vector current is
\be
j_\mu^{\rm CI} \equiv j_\mu^{\rm C} (x) + \frac{r}{2} C_{CVC}\, a \sum_\rho
\partial_\rho \bigl( \overline{\psi}(x) \sigma_{\rho\mu}\psi(x)\bigr)
\, ,
\label{impconserved}
\ee
where $C_{CVC}$ is the improvement coefficient for the conserved
vector current and we define
\be
\partial_\rho \bigl( \overline{\psi}(x) \psi(x)\bigr) \equiv
\overline{\psi}(x) \bigl( \overleftarrow{\nabla}_\rho +
\overrightarrow{\nabla}_\rho \bigr) \psi(x)\, ,
\label{derCVC}
\ee
where the forward and backward derivatives are defined as
\begin{eqnarray*}
\overrightarrow{\nabla}_{\mu}\psi(x) &=&
\frac{1}{2a} \left [
U_{\mu}(x)\, \psi(x+\hat{\mu}) \right . \\  
&&\qquad\qquad
\left . - U^{\dagger}_{\mu}(x-\hat{\mu}) \, \psi(x-\hat{\mu})
\right ] \, ,\\
\overline{\psi}(x)\overleftarrow{\nabla}_{\mu} &=& \frac{1}{2a} \left [
\overline\psi(x+\hat{\mu})\,  U_{\mu}^\dagger(x) \right . \\
&&\qquad\qquad
\left . - \overline\psi(x-\hat{\mu}) \, U_{\mu}(x-\hat{\mu})
\right ]\, .
\end{eqnarray*}

The terms proportional to the Wilson parameter $r$ in
Eq.~(\ref{conserved}) and the four-divergence in
Eq.~(\ref{impconserved}) have their origin in the irrelevant operators
of the fermion action and vanish in the continuum limit.
Non-perturbative improvement is achieved by constructing these terms
with fat-links.  As we have stated, perturbative corrections are small
for fat-links and the use of the tree-level value for $C_{CVC} = 1$
together with small mean-field improvement corrections ensures that
${\cal O}(a)$ artifacts are accurately removed from the vector
current.  This is only possible when the current is constructed with
fat-links.  Otherwise, $C_{CVC}$ needs to be appropriately tuned to
ensure all ${\cal O}(a)$ artifacts are removed.

In order to suppress contributions from excited states, large
Euclidean times are required, both following the source at $t_0$, and
following the current insertion at $t_1$.  Our two-point function
analysis indicates that the ground state is isolated well by $t=14$,
largely due to an excellent selection for the source smearing
parameters.  Therefore the current insertion is performed at $t_1 =
14$.

\subsection{Improved unbiased estimators}

The two and three-point correlation functions are defined as averages
over an infinite ensemble of equilibrium gauge field configurations,
but are approximated by an average over a finite number of
configurations.
To minimise the noise in the results, we exploit the parity of the
correlation functions \cite{Draper:1988xv}
\be
 G(\vec{p'},\vec p, \vec q; \Gamma) 
     = s_P \, G(-\vec{p'},-\vec p,-\vec q; \Gamma), \qquad s_P=\pm 1 ,
\label{parity}
\ee
and calculate them for both $\vec p,\, \vec {p'},\, \vec q$ and $-\vec
p,\, -\vec {p'},\, -\vec q$.
While this requires an extra matrix inversion to determine $\widehat
S(x_2,0;t_1,-\vec q,\mu)$, the ratio of three- to two-point functions
is determined with a substantial reduction in the statistical
uncertainties.  The improvement is better than that obtained by
doubling the number of configurations.

Similarly, the link variables $\{U\}$ and $\{U^*\}$ are gauge field
configurations of equal weight, and therefore we account for both sets
of configurations in calculating the correlation functions
\cite{WM:2002eg}.
With the fermion matrix property
\begin{equation}
 M(\{U^*\}) = \left ( \widetilde C M(\{U\}) \widetilde C^{-1}
                \right)^* ,
\end{equation}
it follows that 
\begin{equation} 
      S(x,0;\{U^*\}) = \left ( \widetilde C S(x,0;\{U\})
                                \widetilde C^{-1} \right )^* ,
\end{equation}
\begin{equation}
\widehat S(x,0;t,\vec q, \mu;\{U^*\}) 
         = \left ( \widetilde C \widehat S(x,0;t,-\vec q,\mu ; \{U\})
                                \widetilde C^{-1} \right )^* ,
\end{equation}
and therefore the correlation functions are purely real provided
\begin{equation}
 \Gamma = s_C \left ( \widetilde C \Gamma \widetilde C^{-1} 
                \right )^* 
\quad {\rm and} \quad s_C = s_P.
\end{equation}
These conditions are satisfied with the selections for $\Gamma$
indicated in Eq.~(\ref{gamma}).
In summary, the inclusion of both $\{U\}$ and $\{U^*\}$
configurations in the calculation of the correlation functions
provides an improved unbiased estimate of the ensemble average
properties incorporating parity symmetry and significantly reducing
statistical fluctuations.

\subsection{Fit regime selection criteria}
\label{fitSel}

In fitting the correlation functions, the covariance-matrix based
$\chi^2$ per degree of freedom ($dof$) plays a central role.

The correlated $\chi^{2}/{\rm dof}$ is given by
\begin{eqnarray}
\frac{\chi^{2}}{\rm dof} &=& \frac{1}{N_{t}-M} \sum_{i=1}^{N_{t}} 
\sum_{j=1}^{N_{t}} \\
&& \left (y(t_{i})-T(t_{i}) \right ) \, C^{-1}(t_{i},t_{j}) \, 
\left ( y(t_{j})-T(t_{j}) \right ) \, , 
\nonumber
\end{eqnarray}
where, $M$ is the number of parameters to be fitted, $N_{t}$ is the
number of time slices considered, $y(t_{i})$ is the configuration
average value of the dependent variable at time $t_{i}$ that is being
fitted to a theoretical value $T(t_{i})$, and $C(t_{i},t_{j})$ is the
covariance matrix.
The elements of the covariance matrix are estimated via the jackknife
method
\begin{eqnarray}
C(t_{i},t_{j}) &=& \frac{N_{c}-1}{N_{c}} \sum_{m=1}^{N_{c}}  \\
&&\left [ \overline{y_{m}}(t_{i}) - \overline{\overline{y}}(t_{i})
\right ] \,  \left [ \overline{y_{m}}(t_{j}) -
\overline{\overline{y}}(t_{j}) \right ] \, , \nonumber \\
&=& \left ( {N_{c}-1} \right ) \times  \\
&&\left \{ \frac{1}{N_{c}} \sum_{m=1}^{N_{c}}
\overline{y_{m}}(t_{i})\, \overline{y_{m}}(t_{j}) -
\overline{\overline{y}}(t_{i})\, \overline{\overline{y}}(t_{j})
\right \} \nonumber
\end{eqnarray} 
where $N_{c}$ is the total number of configurations and
$\overline{y_{m}}(t_{i})$ is the jackknife ensemble average of the
system after removing the ${m_{th}}$ configuration.
$\overline{\overline{y}}(t_{i})$ is the average of all such jackknife
averages, given by
\begin{equation}
\overline{\overline{y}}(t_{i})=\frac{1}{N_{c}}\sum_{m=1}^{N_{c}}\
\overline{y_{m}}(t_{i})\ .
\end{equation}

In the process of selecting the fit regimes, numerous fits are
performed over a variety of start times and a variety of time
durations.
The following criteria are taken into account in selecting the
preferred fit regime:

\begin{enumerate} 
  
    \item The $\chi^2/dof$ is monitored and plays a significant role
  in determining the start time of the fit.
  Values within the range 0.5 to 1.5 are preferred and it is often
  possible to select a regime providing a perfect fit measure of 1.
  Start times for which the $\chi^2/dof$ increases significantly as the
  duration of the regime is increased are discarded.  
  In practice, the $\chi^2/dof$ sets a lower bound for the start time,
  and other criteria may lead to a later start time for the fit.
  
    \item In some cases a monotonic systematic drift can be observed
  in the ratio of correlation functions (\ref{ratio}) which otherwise
  should be constant in time.
  Often the drift is sufficiently small to provide a $\chi^2/dof <
  1.5$.  
  In these few cases, a later start time is selected to ensure that
  sufficient Euclidean time evolution has occurred to accurately
  isolate the ground state, suppressing systematic errors at the
  expense of larger statistical errors.
  
    \item As the quarks become lighter, the spacing between the ground
  and first excited states of the baryon spectrum becomes larger
  \cite{Nstar,WM:2002eg}, due to the more rapid reduction in the mass
  of the lower-lying state.  This provides improved exponential
  suppression of excited-state contaminations.
  Hence, as one approaches the light quark mass regime, a monotonic
  reduction in the starting time-slice of the fit regime may be
  possible.
  
    \item As the quark masses become lighter, the signal is lost to
  noise at earlier times.  
  Hence the final time slice of the fit window is also monotonically
  decreased as the quarks become lighter.  We typically consider fit
  regimes of three to five time slices and preferably the latter when
  the signal is not obviously lost to noise.
  
    \item For quark masses lighter than the strange quark mass, the
  splittings between adjacent quark masses are calculated and fit
  using the same techniques.  
  By considering adjacent splittings, excited-state contributions
  which are less dependent on the quark mass (see item 3 above) are
  suppressed.  In practice, good $\chi^2/dof$ are found one to two
  time slices earlier. 
\end{enumerate}

A precise examination of the environment sensitivity of quark-sector
contributions to baryon electromagnetic properties lies at the core of
this investigation.
For example, the doubly-represented $u$ quark in the proton is to be
compared with the doubly-represented $u$ quark in $\Sigma^+$; the
singly-represented $u$ quark in the neutron with the $u$ quark in
$\Xi^0$.  Similarly, it is interesting to compare the strange and
light quark sectors of $\Xi^-$ with those of $\Lambda$.  Conventional
models reverse the ordering of the observed magnetic moments.
After the consideration of the preceding criteria, the fit regimes are
unified for each of the quark sector contributions wherever
possible.
This comparison is done for each value of $\kappa$ governing the quark
mass, reducing systematic errors associated with choosing different
time-fitting regimes for similar quantities.
For example, for the case of the doubly-represented $u$ quark in the
proton and $\Sigma^+$, it is possible to equate the fit regimes for
all but the two lightest quark masses where the $\chi^2/dof$ insists
on different fit regimes.


\section{CORRELATION FUNCTION ANALYSIS}
\label{sec:corrFun}

The following calculations are performed with $\vec p=0,\,\,\vec{p'}=
\vec q = | \vec q | \, \widehat x$ at $ q_x\, a = 2 \pi / L_x$ with
$L_x = 20$, the minimum non-zero momentum available on our lattice.
We introduce $Q^2 = -q^2$, as $q^2$ is negative (space-like).
While $Q^2$ is dependent on the mass of the baryon, we find this mass
dependence to be small.  Indeed all form factors may be regarded as
being calculated at $Q^2 = 0.227 \pm 0.002\ {\rm GeV}^2$ where the
error is dominated by the mass dependence of the target baryon.
Where a spatial direction of the electromagnetic current is required,
it is chosen to be the $z$-direction.
Electric and magnetic form factors are extracted from our correlation
functions as described in Eqs.~(\ref{GEratio}) and (\ref{GMratio}).

\subsection{Baryon masses}

\begin{figure}[tbp!]
\begin{center}
  {\includegraphics[height=\hsize,angle=90]{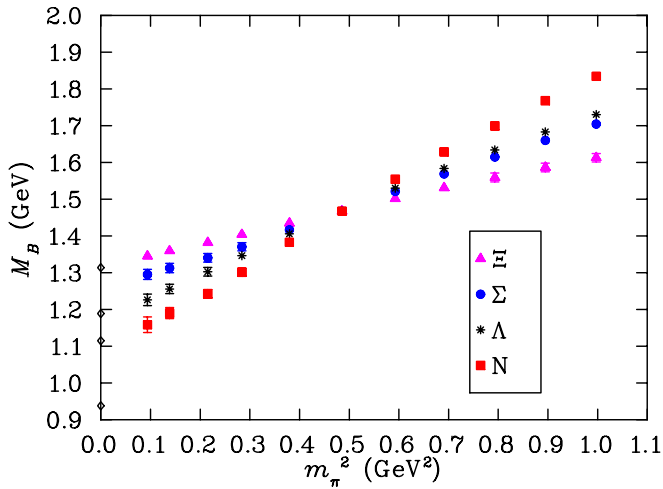}}
\end{center}
\vspace*{-0.5cm}
\caption{Masses of the octet baryons.  The SU(3) flavour limit is
  evident.  Points on the y-axis indicate the experimentally measured
  masses for reference.}
\label{octetmass}
\end{figure}

The masses of the baryon octet are plotted against $m_\pi^2$ in
Fig.~\ref{octetmass} and are tabulated in
Table~\ref{tab:baryonmasses}.
We observe the $SU(3)_f$ limit at our sixth quark mass.
The mass splitting between $\Sigma$ and $\Lambda$ at the lowest pion
mass ($m_\pi=0.3064 \pm 0.0072\:\mathrm{GeV}$) on our lattice is $69 \pm
2\ \mathrm{MeV}$ which is only slightly smaller than the experimentally
measured splitting of $76\ \mathrm{MeV}$.  Hence the generic features
of the baryon-octet mass spectrum is reproduced well in our quenched
simulation.

\subsection{Form factor correlators}
\label{ffCorr}

In general, the baryon form factors are calculated on a quark-sector
by quark-sector basis with each sector normalised to the
contribution of a single quark with unit charge.
Hence to calculate the corresponding baryon property, each quark
sector contribution should be multiplied by the appropriate charge and
quark number.  
Under such a scheme for a generic form factor $f$, the proton form
factor, $f_p$, is obtained from the $u$- and $d$-quark sectors
normalised for a single quark of unit charge via
\begin{equation}
{f_p} = 2 \times \frac{2}{3} \times {f_u} + 1
\times\left(-\frac{1}{3}\right) \times {f_d}\,.
\label{reconsEg}
\end{equation}

\begin{figure}[tbp!]
\begin{center}
 {\includegraphics[height=\hsize,angle=90]{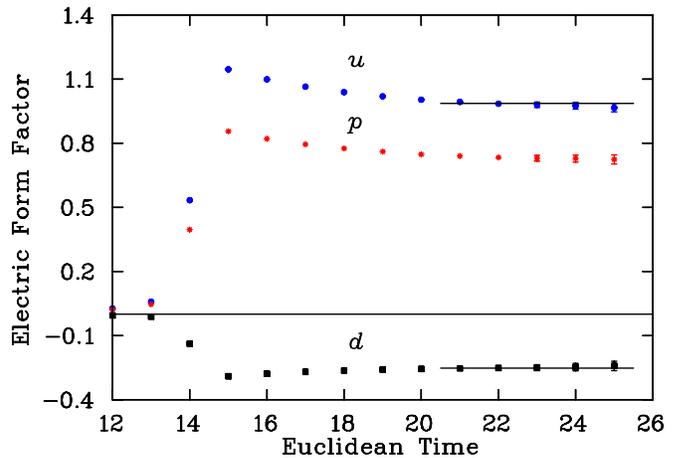}}
\end{center}
\caption{Electric form factor of the proton and its quark sectors
  (including charge and quark number factors) at $Q^2=0.227(2)$
  $\mathrm{GeV}^2$ as a function of Euclidean time ($t_2$) for
  ${m}_{\pi}^{2}=0.4854\ \mathrm{GeV}^2$, the
  $SU(3)$-flavour limit.  The lines indicate the fitting windows and
  the best fit value.}
\vspace{-0.2cm}
\label{cfeK3N}
\end{figure}

The electric form factor of the proton and contributions from the $u$-
and $d$-quark sectors are plotted in Fig.~\ref{cfeK3N} as a function
of Euclidean time at the $SU(3)$-flavour limit.  Here, charge and quark
number factors have been included such that the proton result is
simply the sum of the illustrated quark sectors.
The lines indicate the time slices  selected for the
fit using the considerations of Sec.~\ref{fitSel}.

We find that substantial Euclidean time evolution is required
following the current insertion to obtain acceptable values of the
$\chi^2/dof$; in this case seven time slices following the current
insertion at $t_1 = 14$.

\begin{figure}[tbp!]
\begin{center}
 {\includegraphics[height=\hsize,angle=90]{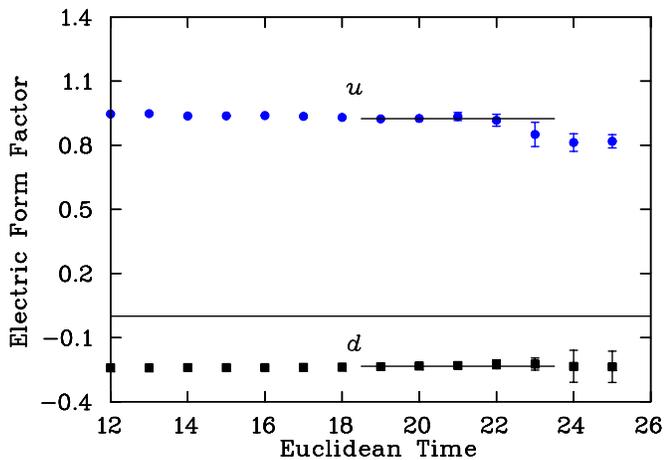}}
\end{center}
\caption{Quark sector contributions (including charge and quark number
  factors) to the electric form factor of the proton at $Q^2=0.227(2)$
  $(\mathrm{GeV}^2)$ as a function of Euclidean time, $t_2$, for the
  ninth quark mass where $m_\pi^2 = 0.2153(35)\ {\rm GeV}^2$.  The
  correlator is obtained from the splitting between the ninth and
  eighth quark mass states.  The lines indicate the fitting
  windows and the best fit value.}
\vspace{-0.2cm}
\label{cfeK6NSp}
\end{figure}

For light quark masses lighter than the strange quark mass, we fit the
change in the form factor ratios of Eq.~(\ref{ratio}) from one quark
mass to the next and add this to the previous result at the heavier
quark mass.
Figure~\ref{cfeK6NSp} shows the quark sector contributions (including
charge and quark number factors) to the electric form factor of the
proton at $Q^2=0.227(2)$ $(\mathrm{GeV}^2)$ as a function of Euclidean
time, $t_2$, for the ninth quark mass where $m_\pi^2 = 0.2153(35)\
{\rm GeV}^2$.  The correlator is obtained from the splitting between
the ninth and eighth quark mass states.  The improvement of the
plateau is apparent in Fig.~\ref{cfeK6NSp}.  Still substantial
Euclidean time evolution is required to obtain an acceptable
$\chi^2/dof$.  The onset of noise at this lighter quark mass is
particularly apparent at time slice 24 for the $d$ sector.
Tables~\ref{tab:effN} to \ref{tab:effX} list the electric form factors
for all the octet baryons at the quark level for the eleven quark masses
considered.
In the tables, the selected time frame, the fit value and the
associated $\chi^2/dof$ are indicated.

\begin{table*}[tbp]
\caption{Quark sector contributions to the electric form factors of
  the nucleon at $Q^2 = 0.227(2)\ {\rm GeV}^2$.  Sector contributions
  are for a single quark having unit charge. The fit windows are
  selected using the criteria outlined in Sec.~\ref{fitSel}.}
\label{tab:effN}
\begin{ruledtabular}
\begin{tabular}{ccccccc}
\noalign{\smallskip}
$ m_\pi^2$  &\multicolumn{3}{c}{$ u_p$}  & \multicolumn{3}{c}{ $ d_p$ }\\
\noalign{\smallskip}
$({\rm GeV}^2)$ & fit value & fit window & $\chi^2/{dof}$ &fit value & fit window & $\chi^2/{dof}$   \\
\hline
\noalign{\smallskip}
$0.9972(55)$ & $0.798(5)$ & $21-25$  & $1.02$  &  $0.805(4)$   & $21-25$  & $2.36$ \\
$0.8947(54)$ & $0.789(5)$ & $21-25$  & $0.89$  &  $0.796(5)$   & $21-25$  & $2.53$ \\
$0.7931(53)$ & $0.779(6)$ & $21-25$  & $0.64$  &  $0.788(5)$   & $21-25$  & $2.21$ \\
$0.6910(35)$ & $0.768(6)$ & $21-25$  & $0.86$  &  $0.780(5)$   & $21-25$  & $1.57$ \\
$0.5925(33)$ & $0.756(7)$ & $21-25$  & $0.80$  & $0.769(6)$    & $21-25$  &  $1.42$ \\
$0.4854(31)$ &  $0.740(9)$ & $21-25$  & $0.62$  &  $0.755(9)$  & $21-25$  & $1.19$  \\
$0.3795(31)$ &  $0.725(10)$ & $19-23$ & $1.23$  &  $0.741(11)$  & $19-23$  & $0.78$ \\
$0.2839(33)$ & $0.708(12)$ & $19-23$ & $1.37$  &  $0.723(14)$  & $19-23$  & $1.31$ \\
$0.2153(35)$ & $0.693(15)$ & $19-23$ & $0.82$  &  $0.700(20)$  & $19-23$  & $1.23$ \\
$0.1384(43)$ & $0.682(17)$ & $16-20$ & $1.02$  &  $0.678(25)$  & $16-20$  & $0.89$ \\
$0.0939(44)$ &  $0.666(25)$ & $16-19$ & $1.47$  &  $0.644(38)$  & $16-19$  & $1.28$ \\
\end{tabular}
\end{ruledtabular}
\vspace{-3pt}
\end{table*}

\begin{table*}[tbp]
\caption{Quark sector contributions to the electric form factors of
  $\Sigma$ baryons at $Q^2 = 0.227(2)\ {\rm GeV}^2$.  Sector contributions
  are for a single quark having unit charge. The fit windows are
  selected using the criteria outlined in Sec.~\ref{fitSel}.}
\label{tab:effS}
\begin{ruledtabular}
\begin{tabular}{ccccccc}
\noalign{\smallskip}
$ m_\pi^2$  &\multicolumn{3}{c}{$ u_\Sigma$ or $ d_\Sigma$}  & \multicolumn{3}{c}{ $ s_\Sigma$ }\\
\noalign{\smallskip}
$({\rm GeV}^2)$ & fit value & fit window & $\chi^2/{dof}$ &fit value & fit window & $\chi^2/{dof}$   \\
\hline
\noalign{\smallskip}
$0.9972(55)$ & $0.793(6)$ & $21-25$  & $0.91$  &  $0.759(6)$   & $21-25$  & $1.70$ \\
$0.8947(54)$ & $0.785(7)$ & $21-25$  & $0.86$  &  $0.758(7)$   & $21-25$  & $1.71$ \\
$0.7931(53)$ & $0.776(7)$ & $21-25$  & $0.66$  &  $0.757(8)$   & $21-25$  & $1.59$ \\
$0.6910(35)$ &  $0.766(6)$ & $21-25$  & $1.00$  &  $0.757(6)$  & $21-25$  & $1.40$ \\
$0.5925(33)$ &  $0.755(7)$ & $21-25$  & $0.90$  & $0.756(7)$   & $21-25$  & $1.36$ \\
$0.4854(31)$ & $0.740(9)$ & $21-25$  & $0.62$  &  $0.755(9)$  & $21-25$  & $1.19$  \\
$0.3795(31)$ & $0.726(10)$ & $19-23$ & $1.46$  &  $0.754(10)$  & $19-23$  & $0.37$ \\
$0.2839(33)$ & $0.711(12)$ & $19-23$ & $1.78$  &  $0.753(11)$  & $19-23$  & $0.58$ \\
$0.2153(35)$ & $0.700(14)$ & $19-23$ & $0.73$  &  $0.752(13)$  & $19-23$  & $0.39$ \\
$0.1384(43)$ & $0.680(18)$ & $19-21$ & $0.73$  &  $0.754(17)$  & $19-21$  & $0.18$ \\
$0.0939(44)$ & $0.670(23)$ & $19-23$ & $1.30$  &  $0.750(26)$  & $19-21$  & $1.40$ \\
\end{tabular}
\end{ruledtabular}
\vspace{-3pt}
\end{table*}

\begin{table*}[tbp]
\caption{Quark sector contributions to the electric form factors of
  $\Lambda$ at $Q^2 = 0.227(2)\ {\rm GeV}^2$.  Sector contributions
  are for a single quark having unit charge. The fit windows are
  selected using the criteria outlined in Sec.~\ref{fitSel}.}
\label{tab:effL}
\begin{ruledtabular}
\begin{tabular}{ccccccc}
\noalign{\smallskip}
$ m_\pi^2$  &\multicolumn{3}{c}{$ u_\Lambda$ or $ d_\Lambda$}  & \multicolumn{3}{c}{ $ s_\Lambda$ }\\
\noalign{\smallskip}
$({\rm GeV}^2)$ & fit value & fit window & $\chi^2/{dof}$ &fit value & fit window & $\chi^2/{dof}$   \\
\hline
\noalign{\smallskip}
$0.9972(55)$ &  $0.803(5)$ & $21-25$  & $1.20$  &  $0.745(8)$   & $21-25$  & $0.64$ \\
$0.8947(54)$ &  $0.794(6)$ & $21-25$  & $1.23$  &  $0.744(9)$   & $21-25$  & $0.58$ \\
$0.7931(53)$ &  $0.785(7)$ & $21-25$  & $1.06$  &  $0.744(10)$  & $21-25$  & $0.54$ \\
$0.6910(35)$ &  $0.775(6)$ & $21-25$  & $1.17$  &  $0.738(8)$   & $21-25$  & $0.55$ \\
$0.5925(33)$ &  $0.765(7)$ & $21-25$  & $1.12$ &   $0.737(9)$   & $21-25$  & $0.48$ \\
$0.4854(31)$ &  $0.750(8)$ & $21-25$  & $1.02$ &  $0.735(10)$   & $21-25$  & $0.49$  \\
$0.3795(31)$ &  $0.736(9)$ & $19-23$  & $0.94$ &  $0.734(11)$  & $19-23$  & $0.88$ \\
$0.2839(33)$ &  $0.720(11)$ & $19-23$ & $1.17$ &  $0.730(12)$  & $19-23$  & $1.05$ \\
$0.2153(35)$ &  $0.704(13)$ & $19-23$ & $1.23$ &  $0.727(13)$  & $19-23$  & $0.55$ \\
$0.1384(43)$ &  $0.694(13)$ & $16-19$ & $3.79$ &  $0.727(13)$  & $16-17$  & $0.62$ \\
$0.0939(44)$ &  $0.686(13)$ & $16-17$ & $2.43$ &  $0.729(14)$  & $16-17$  & $0.29$ \\
\end{tabular}
\end{ruledtabular}
\vspace{-3pt}
\end{table*}

\begin{table*}[tbp]
\caption{Quark sector contributions to the electric form factors of
  $\Xi$ baryons at $Q^2 = 0.227(2)\ {\rm GeV}^2$.  Sector contributions
  are for a single quark having unit charge. The fit windows are
  selected using the criteria outlined in Sec.~\ref{fitSel}.}
\label{tab:effX}
\begin{ruledtabular}
\begin{tabular}{ccccccc}
\noalign{\smallskip}
$ m_\pi^2$  & \multicolumn{3}{c}{ $ s_\Xi$ } &\multicolumn{3}{c}{$ u_\Xi$ or $ d_\Xi$}  \\
\noalign{\smallskip}
$({\rm GeV}^2)$ & fit value & fit window & $\chi^2/{dof}$ &fit value & fit window & $\chi^2/{dof}$   \\
\hline
\noalign{\smallskip}
$0.9972(55)$ & $0.747(9)$ & $21-25$  & $0.34$  &  $0.804(8)$   & $21-25$  & $1.60$ \\
$0.8947(54)$ & $0.747(9)$ & $21-25$  & $0.36$  &  $0.794(8)$   & $21-25$  & $1.53$ \\
$0.7931(53)$ & $0.746(10)$ & $21-25$  & $0.37$ &  $0.785(9)$ & $21-25$  & $1.52$ \\
$0.6910(35)$ & $0.742(8)$ & $21-25$  & $0.54$ &  $0.778(7)$ & $21-25$  & $1.38$ \\
$0.5925(33)$ & $0.741(8)$ & $21-25$  & $0.55$ & $0.768(8)$ & $21-25$  &  $1.24$ \\ 
$0.4854(31)$ & $0.740(9)$ & $21-25$  & $0.62$ &  $0.755(9)$ & $21-25$  & $1.19$  \\
$0.3795(31)$ & $0.739(9)$ & $19-23$  & $0.70$ &  $0.740(10)$  & $21-25$  & $1.45$ \\
$0.2839(33)$ &  $0.738(10)$ & $19-23$   & $1.18$ &  $0.723(13)$  & $21-25$  & $1.22$ \\
$0.2153(35)$ &  $0.736(10)$ & $19-23$ & $1.42$ &  $0.709(16)$  & $21-25$  & $0.81$ \\
$0.1384(43)$ & $0.733(10)$ & $19-23$ & $0.52$ &  $0.690(19)$  & $20-23$  & $0.71$ \\
$0.0939(44)$ & $0.725(11)$ & $19-23$ & $1.21$ &  $0.672(22)$  & $20-23$  & $0.59$ \\
\end{tabular}
\end{ruledtabular}
\vspace{-3pt}
\end{table*}

\begin{figure}[tbp!]
\begin{center}
 {\includegraphics[height=\hsize,angle=90]{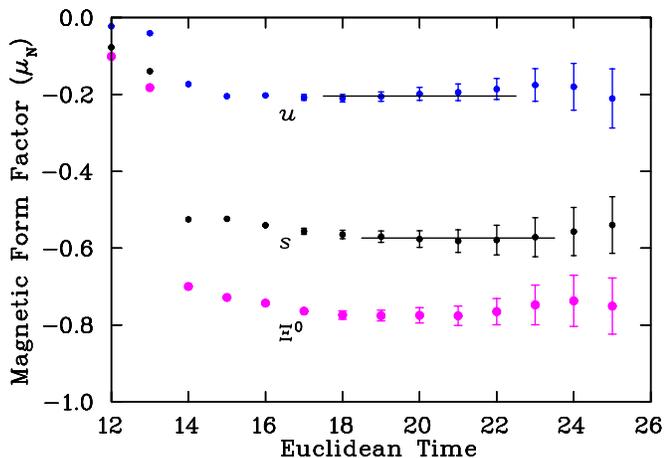}}
\end{center}
\vspace*{-0.5cm}
\caption{Magnetic form factors of $\Xi^0$ and its quark sectors
  (including charge and quark number factors) at $Q^2=0.227(2)$
  $\mathrm{GeV}^2$ as a function of Euclidean time ($t_2$) for
  ${m}_{\pi}^{2}=0.4854\ \mathrm{GeV}^2$, the $SU(3)$-flavour
  limit.  The lines indicate the fitting windows and the best fit
  value.}
\vspace{-0.2cm}
\label{cfmK3X}
\end{figure}

Turning now to the magnetic form factors, Fig.~\ref{cfmK3X} shows the
magnetic form factor of $\Xi^0$ and its quark sectors (including
charge and quark number factors) as a function of Euclidean time at
the $SU(3)$-flavour limit.  Preferred fit windows following from the
criteria of Sec.~\ref{fitSel} and best fit values are indicated.

Here the conversion from the natural magneton, $e/(2\, m_B)$, where
the mass of the baryon under investigation appears, to the nuclear
magneton, $e/(2\, m_N)$, where the physical nucleon mass appears, has
been done by multiplying the lattice form factor results by the ratio
$m_N/m_B$.  In this way the form factors are presented in terms of a
constant unit; {\it i.e.} the nuclear magneton.

The negative contribution of the ${u}$ quark to the total form
factor indicates that its spin is on average opposite to that of the
doubly represented ${s}$ quarks.  This, as well the relative
magnitude of the contributions, is in qualitative agreement with simple
constituent quark models based on $SU(6)$ spin-flavour symmetry.

\begin{figure}[tbp!]
\begin{center}
 {\includegraphics[height=\hsize,angle=90]{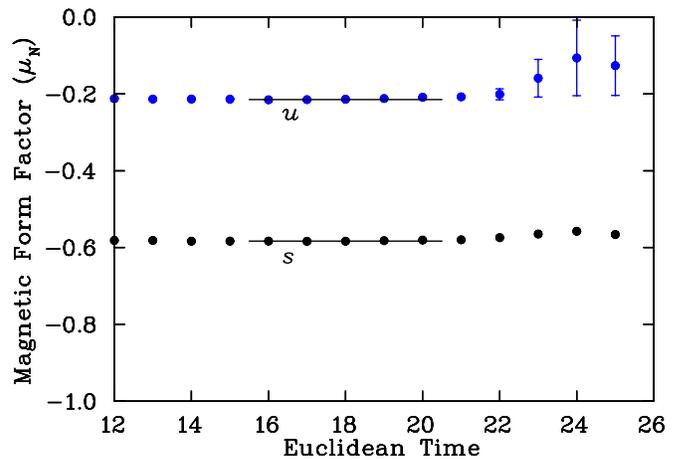}}
\end{center}
\vspace*{-0.5cm}
\caption{Quark sector contributions (including charge and quark number
  factors) to the magnetic form factor of $\Xi^0$ at $Q^2=0.227(2)$
  $(\mathrm{GeV}^2)$ as a function of Euclidean time, $t_2$, for the
  ninth quark mass where $m_\pi^2 = 0.2153(35)\ {\rm GeV}^2$.  The
  correlator is obtained from the splitting between the ninth and
  eighth quark mass states.  The lines indicate the fitting
  windows and the best fit value.}
\vspace{-0.2cm}
\label{cfmK6XSp}
\end{figure}

In Fig.~\ref{cfmK6XSp} we present the Euclidean time dependence of the
the magnetic form factors of $\Xi^0$ calculated at the ninth quark
mass where $m_\pi^2 = 0.2153(35)\ {\rm GeV}^2$.  Again the early onset
of acceptable  plateau behaviour is apparent here.

Results for the quark-sector contributions to the magnetic form
factors of octet baryons are summarised in Tables~\ref{tab:mffN} to
\ref{tab:mffX}.

\begin{table*}[tbp]
\caption{Quark sector contributions to the magnetic form factors of
  the nucleon at $Q^2 = 0.227(2)\ {\rm GeV}^2$.  Sector contributions
  are for a single quark having unit charge. The fit windows are
  selected using the criteria outlined in Sec.~\ref{fitSel}.}
\label{tab:mffN}
\begin{ruledtabular}
\begin{tabular}{ccccccc}
\noalign{\smallskip}
$ m_\pi^2\ ({\rm GeV}^2)$  &\multicolumn{3}{c}{$ u_p\
  (\mu_N)$}  & \multicolumn{3}{c}{ $ d_p\ (\mu_N)$ }\\
\noalign{\smallskip}
 & fit value & fit window & $\chi^2/{dof}$ &fit value & fit window & $\chi^2/{dof}$   \\
\hline
\noalign{\smallskip}
$0.9972(55)$ &  $0.765(12)$ & $19-23$  & $1.78$  &  $-0.295(7)$   & $18-22$  & $0.61$ \\
$0.8947(54)$ &  $0.785(14)$ & $19-23$  & $1.33$  &  $-0.298(8)$   & $18-22$  & $0.51$ \\
$0.7931(53)$ &  $0.804(16)$ & $19-23$  & $1.01$  &  $-0.301(9)$  & $18-22$  & $0.43$ \\
$0.6931(51)$ & $0.817(13)$ & $19-23$   & $1.00$  &  $-0.301(8)$  & $18-22$  & $0.91$ \\
$0.5944(51)$ & $0.838(15)$ & $19-23$   & $0.73$  &  $-0.304(10)$  & $18-22$  & $0.79$ \\
$0.4869(50)$ & $0.861(20)$ & $19-23$   & $0.64$  &  $-0.306(12)$  & $18-22$  & $0.86$ \\
$0.3795(31)$ & $0.893(24)$ & $17-21$   & $0.14$ &  $-0.314(14)$  & $16-18$  & $1.64$ \\
$0.2839(33)$ & $0.932(31)$ & $17-21$   & $0.14$ &  $-0.313(19)$  & $16-19$  & $1.24$ \\
$0.2153(35)$ & $0.967(42)$ & $17-21$   & $0.61$ &  $-0.313(31)$  & $16-20$  & $0.53$ \\
$0.1384(43)$ & $1.034(52)$ & $16-20$   & $1.12$ &  $-0.309(40)$  & $15-19$  & $0.49$ \\
$0.0939(44)$ & $1.024(72)$ & $15-17$   & $0.82$ &  $-0.335(54)$  & $15-17$  & $2.44$ \\
\end{tabular}
\end{ruledtabular}
\vspace{-3pt}
\end{table*}

\begin{table*}[tbp]
\caption{Quark sector contributions to the magnetic form factors of
  $\Sigma$ baryons at $Q^2 = 0.227(2)\ {\rm GeV}^2$.  Sector contributions
  are for a single quark having unit charge. The fit windows are
  selected using the criteria outlined in Sec.~\ref{fitSel}.}
\label{tab:mffS}
\begin{ruledtabular}
\begin{tabular}{ccccccc}
\noalign{\smallskip}
$ m_\pi^2\ ({\rm GeV}^2)$  
&\multicolumn{3}{c}{$u_\Sigma\ {\rm or}\ d_\Sigma\ (\mu_N)$} 
&\multicolumn{3}{c}{$ s_\Sigma\ (\mu_N)$}
\\
\noalign{\smallskip}
 & fit value & fit window & $\chi^2/{dof}$ &fit value & fit window & $\chi^2/{dof}$   \\
\hline
\noalign{\smallskip}
\noalign{\smallskip}
$0.9972(55)$ &  $0.775(14)$ & $19-23$  & $1.22$   &  $-0.316(10)$ & $18-22$  & $0.73$ \\
$0.8947(54)$ &  $0.793(16)$ & $19-23$  & $1.10$   &  $-0.314(11)$ & $18-22$  & $0.63$ \\
$0.7931(53)$ &  $0.810(18)$ & $19-23$  & $0.96$   &  $-0.312(11)$ & $18-22$  & $0.55$ \\
$0.6910(35)$ &  $0.821(14)$ & $19-23$  & $0.98$   &  $-0.309(9)$  & $18-22$  & $0.94$ \\
$0.5925(33)$ &  $0.840(16)$ & $19-23$  & $0.82$   &  $-0.308(10)$ & $18-22$  & $0.86$ \\
$0.4854(31)$ &  $0.861(20)$ & $19-23$  & $0.64$   &  $-0.306(12)$ & $18-22$  & $0.86$  \\
$0.3795(31)$ &  $0.886(23)$ & $17-21$  & $0.19$   &  $-0.308(13)$ & $16-18$  & $0.42$ \\
$0.2839(33)$ &  $0.914(27)$ & $17-21$  & $0.26$   &  $-0.310(15)$ & $16-18$  & $0.03$ \\
$0.2153(35)$ &  $0.941(32)$ & $17-21$  & $0.61$   &  $-0.317(19)$ & $16-18$  & $0.04$ \\
$0.1384(43)$ &  $0.964(33)$ & $15-20$  & $1.57$   &  $-0.317(24)$ & $16-18$  & $0.92$ \\
$0.0939(44)$ &  $0.969(41)$ & $15-17$ & $0.11$    &  $-0.322(28)$ & $16-18$  & $1.08$ \\
\end{tabular}
\end{ruledtabular}
\vspace{-3pt}
\end{table*}

\begin{table*}[tbp]
\caption{Quark sector contributions to the magnetic form factors of
  $\Lambda$ at $Q^2 = 0.227(2)\ {\rm GeV}^2$.  Sector contributions
  are for a single quark having unit charge. The fit windows are
  selected using the criteria outlined in Sec.~\ref{fitSel}.}
\label{tab:mffL}
\begin{ruledtabular}
\begin{tabular}{ccccccc}
\noalign{\smallskip}
$ m_\pi^2\ ({\rm GeV}^2)$  
&\multicolumn{3}{c}{$u_\Lambda\ {\rm or}\ d_\Lambda\ (\mu_N)$} 
&\multicolumn{3}{c}{$ s_\Lambda\ (\mu_N)$}
\\
\noalign{\smallskip}
 & fit value & fit window & $\chi^2/{dof}$ &fit value & fit window & $\chi^2/{dof}$   \\
\hline
\noalign{\smallskip}
\noalign{\smallskip}
$0.9972(55)$ &  $0.069(9)$ & $19-23$  & $1.26$   &  $1.200(22)$ & $18-22$  & $1.38$ \\
$0.8947(54)$ &  $0.072(9)$ & $19-23$  & $1.20$   &  $1.210(22)$ & $18-22$  & $1.17$ \\
$0.7931(53)$ &  $0.075(10)$ & $19-23$ & $1.07$   &  $1.214(24)$ & $18-22$  & $0.99$ \\
$0.6910(35)$ & $0.079(8)$ & $19-23$  & $1.08$    &  $1.217(17)$ & $18-22$  & $0.95$ \\
$0.5925(33)$ & $0.083(9)$ & $19-23$  & $1.00$    &  $1.228(18)$ & $18-22$  &  $0.80$ \\
$0.4854(31)$ &  $0.087(11)$ & $19-23$ & $0.88$  &  $1.241(21)$ & $18-22$  & $0.75$  \\ 
$0.3795(31)$ & $0.091(12)$ & $17-21$  & $0.55$  &  $1.259(22)$  & $16-20$  & $0.59$ \\
$0.2839(33)$ & $0.097(13)$ & $17-21$   & $0.29$ &  $1.277(25)$  & $16-20$  & $0.42$ \\
$0.2153(35)$ & $0.099(17)$ & $17-21$ & $0.36$ &    $1.293(28)$  & $16-20$  & $0.42$ \\
$0.1384(43)$ &  $0.105(18)$ & $15-16$ & $1.79$ &  $1.308(33)$  & $16-18$  & $0.94$ \\
$0.0939(44)$ &  $0.105(22)$ & $15-16$ & $0.19$ &  $1.315(37)$  & $15-16$  & $1.98$ \\
\end{tabular}
\end{ruledtabular}
\vspace{-3pt}
\end{table*}

\begin{table*}[tbp]
\caption{Quark sector contributions to the magnetic form factors of
  $\Xi$ baryons at $Q^2 = 0.227(2)\ {\rm GeV}^2$.  Sector contributions
  are for a single quark having unit charge. The fit windows are
  selected using the criteria outlined in Sec.~\ref{fitSel}.}
\label{tab:mffX}
\begin{ruledtabular}
\begin{tabular}{ccccccc}
\noalign{\smallskip}
$ m_\pi^2\ ({\rm GeV}^2)$  &\multicolumn{3}{c}{$ s_\Xi\
  (\mu_N)$}  & \multicolumn{3}{c}{ $u_\Xi\ {\rm or}\ d_\Xi\ (\mu_N)$ }\\
\noalign{\smallskip}
 & fit value & fit window & $\chi^2/{dof}$ &fit value & fit window & $\chi^2/{dof}$   \\
\hline
\noalign{\smallskip}
$0.9972(55)$ &  $0.846(27)$ & $19-23$  & $0.60$  &  $-0.290(10)$   & $18-22$  & $0.19$ \\
$0.8947(54)$ &  $0.849(27)$ & $19-23$  & $0.67$  &  $-0.294(11)$   & $18-22$  & $0.21$ \\
$0.7931(53)$ &  $0.851(28)$ & $19-23$  & $0.75$  &  $-0.299(12)$  & $18-22$  & $0.21$ \\
$0.6931(51)$ &  $0.855(19)$ & $19-23$ & $0.41$  &  $-0.300(10)$  & $18-22$  & $0.59$ \\
$0.5944(51)$ &  $0.858(19)$ & $19-23$ & $0.49$  &  $-0.303(11)$  & $18-22$  & $0.68$ \\
$0.4869(50)$ &  $0.861(20)$ & $19-23$ & $0.64$  &  $-0.306(12)$  & $18-22$  & $0.86$ \\
$0.3795(31)$ &  $0.866(21)$ & $16-20$ & $0.84$ &  $-0.311(13)$  & $16-20$  & $1.77$ \\
$0.2839(33)$ &  $0.871(21)$ & $16-20$ & $0.78$ &  $-0.316(14)$  & $16-20$  & $1.66$ \\
$0.2153(35)$ &  $0.875(22)$ & $16-20$ & $1.04$ &  $-0.322(15)$  & $16-20$  & $0.77$ \\
$0.1384(43)$ & $0.879(22)$ & $16-20$ & $1.16$ &  $-0.328(17)$  & $15-19$  & $0.86$ \\
$0.0939(44)$ & $0.879(23)$ & $16-20$ & $0.55$ &  $-0.333(18)$  & $15-19$  & $1.00$ \\
\end{tabular}
\end{ruledtabular}
\vspace{-3pt}
\end{table*}


\section{DISCUSSION OF RESULTS}
\label{sec:Results}

\subsection {Charge radii}

To make contact with the extensive phenomenology of the field, our
results for the electric form factors are expressed in terms of charge
radii.  
It is well known that the experimentally measured electric (and
magnetic) form factor of the proton is described well by a dipole
ansatz at small $Q^2$
\begin{equation}
\quad {\cal G}_E(Q^2) = { {\cal G}_E(0) \over 
             \left (1 + Q^2 / m^2 \right )^2 } \, ; \quad
      Q^2 \ge 0\, ,
\label{dipole}
\end{equation}
where $m$ characterises the size of the proton.
This behaviour has also been observed in recent lattice calculations
\cite{Gockeler:2003ay} where many momentum transfers have been considered.
Using this observation, together with 
\begin{equation}
 \left\langle r_E^2 \right\rangle = -6\, \frac{d}{dQ^2}\, {\cal
   G}_E(Q^2) \biggm |_{Q^2=0} \ ,
\label{charad}
\end{equation}
we arrive at an expression which allows us to calculate the electric
charge radius of a baryon using our two available values of the Sach's
electric form factor (${\cal G}_E(Q^2_{\rm min}),\ {\cal G}_E(0)$),
namely
\begin{equation}
\frac{\left\langle r_E^2 \right\rangle}{ {\cal G}_E(0)} = \frac{12}{Q^2}
\left ( \sqrt{ \frac{{\cal G}_E(0)}{{\cal G}_E(Q^2) }} - 1 \right )\ .
\label{charad_dipole}
\end{equation}

While Eq.~(\ref{dipole}) is suitable for a charged baryon, alternative
forms must be considered for neutral baryons where ${\cal G}_E(0) =
0$.

However, we have direct access to the charge distributions of the
individual quark sectors, a subject receiving tremendous experimental
attention in the search for the role of hidden flavour in baryon
structure.  
In this case Eq.~(\ref{charad_dipole}) may be applied to each quark sector
providing an opportunity to determine the charge radii on a sector by
sector basis.

For neutral baryons it becomes a simple matter to construct the charge
radii by first calculating the charge radii for each quark sector.
These quark sectors are then combined using the appropriate charge and
quark number factors as described in Sec.~\ref{ffCorr} to obtain the
total baryon charge radii.  Indeed, all baryon charge radii, including
the charged states, are calculated in this manner.

Tables~\ref{tab:eradN} to \ref{tab:eradX} provide the electric charge
radii of the octet baryons and their quark-sector contributions
normalised to the case of single quarks with unit charge.

\begin{table*}[tbp]
\caption{Nucleon electric charge radii squared.  Quark-sector
  contributions are indicated for single quarks having unit charge.
  Baryon charge states are also summarised.  Values for $m_\pi^2$ and
  $\langle r^2 \rangle$ are in units of ${\rm GeV}^2$ and ${\rm fm}^2$
  respectively.}
\label{tab:eradN}
\begin{ruledtabular}
\begin{tabular}{ccccc}
$ m_\pi^2$ &  $ u_p$ & $ d_p$ & $ p$ & $ n$ \\
\hline
$0.9972(55)$ & $0.243(7)$      &  $0.231(6)$   & $0.247(8)$   & $-0.007(3)$ \\
$0.8947(54)$ & $0.256(8)$      &  $0.245(7)$   & $0.259(9)$   & $-0.007(3)$ \\
$0.7931(53)$ & $0.270(9)$      &  $0.257(8)$   & $0.273(10)$   & $-0.008(4)$ \\ 
$0.6910(35)$ & $0.288(9)$      &  $0.270(8)$   & $0.294(10)$   & $-0.012(4)$ \\
$0.5925(33)$ & $0.307(11)$     &  $0.286(10)$   & $0.314(12)$  & $-0.014(5)$ \\ 
$0.4854(31)$ & $0.332(14)$     &  $0.309(14)$   & $0.340(16)$  & $-0.015(7)$ \\
$0.3795(31)$ & $0.358(17)$     &  $0.332(17)$   & $0.367(19)$  & $-0.017(9)$    \\
$0.2839(33)$ & $0.389(22)$     &  $0.363(24)$   & $0.397(24)$  & $-0.017(12)$  \\
$0.2153(35)$ & $0.416(27)$     &  $0.403(36)$   & $0.420(29)$  & $-0.008(16)$  \\
$0.1384(43)$ & $0.437(31)$     &  $0.445(46)$   & $0.435(32)$  & $ 0.005(21)$  \\
$0.0939(44)$ & $0.467(48)$     &  $0.510(77)$   & $0.452(53)$  & $ 0.029(41)$  \\

\end{tabular}
\end{ruledtabular}

\caption{$\Sigma$ electric charge radii squared.  Quark-sector
  contributions are indicated for single quarks having unit charge.
  Baryon charge states are also summarised where absolute values of
  the $\Sigma^-$ results are reported.  Values for $m_\pi^2$ and
  $\langle r^2 \rangle$ are in units of ${\rm GeV}^2$ and ${\rm fm}^2$
  respectively.}
\label{tab:eradS}
\begin{ruledtabular}
\begin{tabular}{cccccc}
$ m_\pi^2$ &  $u_\Sigma$ or $d_\Sigma$ & $ s_\Sigma$ & $\Sigma^+$ & $\Sigma^0$ & $\Sigma^-$ \\
\hline
$0.9972(55)$ & $0.249(9)$      &  $0.301(10)$   & $0.232(10)$  & $-0.017(3)$    & $0.266(9)$ \\
$0.8947(54)$ & $0.261(10)$     &  $0.302(11)$   & $0.249(12)$  & $-0.013(3)$    & $0.275(9)$\\ 
$0.7931(53)$ & $0.276(11)$     &  $0.304(12)$   & $0.266(15)$  & $-0.010(3)$     & $0.285(11)$\\
$0.6910(35)$ & $0.291(10)$     &  $0.304(10)$   & $0.286(11)$  & $-0.005(2)$    & $0.295(9)$ \\
$0.5925(33)$ & $0.309(11)$     &  $0.306(11)$   & $0.310(13)$  & $ 0.001(3)$    & $0.308(11)$\\ 
$0.4854(31)$ & $0.332(14)$     &  $0.309(14)$   & $0.340(16)$  & $ 0.008(4)$     & $0.324(13)$\\
$0.3795(31)$ & $0.356(16)$     &  $0.310(15)$   & $0.371(19)$  & $0.015(4)$     & $0.341(15)$ \\
$0.2839(33)$ & $0.382(20)$     &  $0.312(18)$   & $0.405(23)$  & $0.023(5)$      & $0.359(18)$  \\
$0.2153(35)$ & $0.401(25)$     &  $0.315(21)$   & $0.430(29)$  & $0.029(6)$      & $0.372(22)$ \\
$0.1384(43)$ & $0.438(33)$     &  $0.311(26)$   & $0.480(38)$  & $ 0.042(8)$     & $0.395(28)$\\
$0.0939(44)$ & $0.456(43)$     &  $0.318(41)$   & $0.503(54)$  & $ 0.046(15)$     & $0.410(37)$  
\end{tabular}
\end{ruledtabular}
\end{table*}

\begin{table*}[tbp]
\caption{$\Lambda$ electric charge radii squared.  Quark-sector
  contributions are indicated for single quarks having unit charge.
  The baryon charge state is also provided.  Values for $m_\pi^2$ and
  $\langle r^2 \rangle$ are in units of ${\rm GeV}^2$ and ${\rm fm}^2$
  respectively.}
\label{tab:eradL}
\begin{ruledtabular}
\begin{tabular}{cccc}
$ m_\pi^2$ &  $u_\Lambda$ or $d_\Lambda$ & $ s_\Lambda$ & $\Lambda^0$  \\
\hline
$0.9972(55)$ & $0.235(8)$  &  $0.322(13)$    & $-0.029(3)$ \\
$0.8947(54)$ & $0.248(9)$  &  $0.323(14)$    & $-0.025(3)$ \\
$0.7931(53)$ & $0.262(10)$  &  $0.324(15)$    & $-0.021(3)$ \\
$0.6910(35)$ & $0.276(9)$  &  $0.335(13)$    & $-0.020(3)$ \\
$0.5925(33)$ & $0.293(10)$   &  $0.336(14)$    & $-0.014(3)$ \\ 
$0.4854(31)$ & $0.316(13)$  &  $0.340(16)$    & $-0.008(4)$ \\
$0.3795(31)$ & $0.340(15)$     &  $0.343(18)$   & $-0.001(4)$    \\
$0.2839(33)$ & $0.367(18)$     &  $0.350(19)$ & $0.006(5)$  \\
$0.2153(35)$ & $0.395(23)$     &  $0.357(22)$ & $0.013(6)$  \\
$0.1384(43)$ & $0.414(24)$     &  $0.356(22)$ & $0.019(6)$  \\
$0.0939(44)$ & $0.428(25)$     &  $0.354(23)$    & $ 0.025(7)$ 
\end{tabular}
\end{ruledtabular}

\caption{$\Xi$ electric charge radii squared.  Quark-sector
  contributions are indicated for single quarks having unit charge.
  Baryon charge states are also summarised where absolute values of
  the $\Xi^-$ results are reported.  Values for $m_\pi^2$ and $\langle
  r^2 \rangle$ are in units of ${\rm GeV}^2$ and ${\rm fm}^2$
  respectively.}
\label{tab:eradX}
\begin{ruledtabular}
\begin{tabular}{ccccc}
$ m_\pi^2$ &  $ s_\Xi$ & $u_\Xi$ or $d_\Xi$ & $\Xi^-$ & $\Xi^0$  \\
\hline
$0.9972(55)$ & $0.319(14)$   &  $0.235(11)$  & $0.290(12)$  & $-0.056(7)$ \\
$0.8947(54)$ & $0.320(15)$   &  $0.249(12)$  & $0.296(13)$  & $-0.047(7)$ \\
$0.7931(53)$ & $0.322(16)$   &  $0.262(13)$  & $0.301(14)$  & $-0.039(8)$ \\
$0.6910(35)$ & $0.329(13)$   &  $0.273(11)$  & $0.310(11)$  & $-0.037(6)$ \\
$0.5925(33)$ & $0.330(14)$   &  $0.288(12)$  & $0.316(12)$  & $-0.028(7)$ \\ 
$0.4854(31)$ & $0.332(14)$   &  $0.309(14)$  & $0.324(13)$  & $-0.015(7)$ \\
$0.3795(31)$ & $0.334(15)$   &  $0.333(17)$  & $0.334(14)$  & $-0.0003(80)$    \\
$0.2839(33)$ & $0.337(15)$   &  $0.361(21)$  & $0.345(16)$  & $0.016(10)$  \\
$0.2153(35)$ & $0.340(16)$   &  $0.385(27)$  & $0.355(18)$  & $0.029(13)$  \\
$0.1384(43)$ & $0.345(16)$   &  $0.419(34)$  & $0.370(19)$  & $0.049(19)$  \\
$0.0939(44)$ & $0.358(18)$   &  $0.451(41)$  & $0.389(23)$  & $0.062(22)$  
\end{tabular}
\end{ruledtabular}
\end{table*}

\subsubsection {Quenched chiral perturbation theory}

The effective field theory formalism of quenched chiral perturbation
theory (\QchiPT) predicts significant contributions to the charge
radii which have their origin in virtual meson-baryon loop
transitions.  These loops give rise to contributions which have a
non-analytic dependence on the quark mass or squared pion mass.  While
the absence of sea-quark loops generally acts to suppress the
magnitude of the coefficients of these terms (and occasionally the
sign is reversed), there are several channels in which these
contributions remain significant.

The leading non-analytic (LNA) and next-to-leading non-analytic (NLNA)
behaviour of charge distribution radii in full QCD are
\begin{eqnarray}
\langle r_E^2\rangle &\!\!=\!\!& \frac{1}{16\,\pi^2\,f^2} 
\sum_{i} \, \biggl [
5\, \beta_i\, \log \left
  (\frac{m_{i}^2}{\mu^2} \right ) - 10\, \beta^{\prime}_i\, 
{\cal G}(m_i, \Delta, \mu) \nonumber \\
      &   & + c_0 + c_2\, m_i^2 + c_4\, m_i^4 \ldots \biggr ]\ .
\label{charad_QCPT}
\end{eqnarray}
Here the sum over $i$ includes the $\pi$ and $K$ pseudoscalar mesons.
The contributions of the various charge states of these mesons are
contained in the coefficients $\beta$ and $\beta'$, reflecting electric
charge and SU(3) axial couplings, $D$, $F$ and $C$.  In quenching the
theory, the coefficients $\beta$ and $\beta'$ are modified to reflect
the absence of sea-quark loops.

The first term arises from octet baryon to octet-baryon -- meson
transitions.  Thus charge radii are characterised by a logarithmic
divergence \cite{Leinweber:1992hj} in the chiral limit ($m_\pi^2
\rightarrow 0$).  In this simple form, the mass splittings between
baryon octet members is neglected.

The second term of Eq.~(\ref{charad_QCPT}) arises from octet baryon to
decuplet-baryon -- meson transitions.  As the splitting between the
baryon octet and decuplet does not vanish in the chiral limit, the
mass splitting, $\Delta = M_\Delta - M_N$, between the nucleon and
$\Delta$ for example, plays an important role.  The function ${\cal
G}(m_i, \Delta, \mu)$ is
\begin{eqnarray}
\lefteqn{
{\cal G}(m, \Delta, \mu)  =  \log \left (\frac{m^2_{i}}{\mu^2} \right
) - }\nonumber \\
  &   & \frac {\Delta}{\sqrt{\Delta^2 -
    m_i^2}}\log \frac {\Delta - \sqrt{\Delta^2 - m_i^2 + i\epsilon}}
        {\Delta + \sqrt{\Delta^2 -m_i^2 +i\epsilon}}\ .
\label{secondTerm}
\end{eqnarray}

\begin{figure}[tbp!]
\begin{center}
 {\includegraphics[height=\hsize,angle=90]{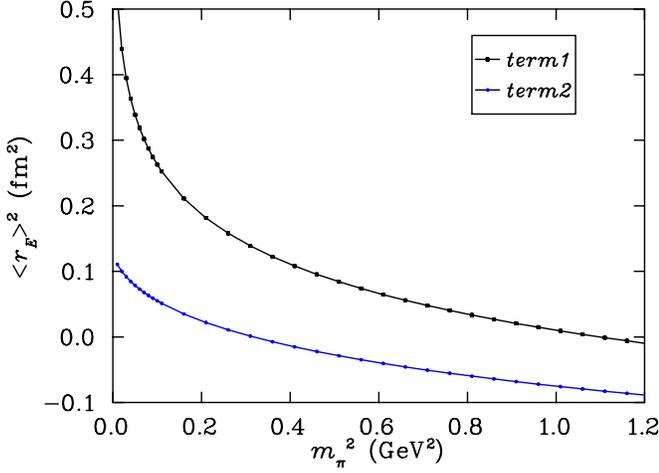}}
\end{center}
\vspace*{-0.5cm}
\caption{The leading (upper curve) and next-to-leading (lower curve)
  non-analytic contributions to the charge radius of the proton as
  given by quenched chiral perturbation theory in
  Eq.~(\ref{charad_QCPT}).  }
\label{charadfig}
\end{figure}

As the tadpole graph contributing to the LNA term of charge radii in
full QCD vanishes in quenched QCD \cite{Arndt:2004}, the coefficients
$\beta$ and $\beta'$ for charge radii are identical to those for
magnetic moments in quenched QCD
\cite{Leinweber:2002qb,Savage:2001dy}.
Figure~\ref{charadfig} displays the non-analytic contributions from
Q$\chi$PT as given in Eq.~(\ref{charad_QCPT}), plotted for the sample
case of the proton.  In this case, the values of $\beta$ and $\beta
^{\prime}$ are $-\frac{4}{3}D^2$ and $-\frac{1}{6}C^2$ respectively
\cite{Leinweber:2002qb,Savage:2001dy}.
Here the axial couplings $D$ and $C$ are related by $C=2D$ and $D$ is
taken as $0.76$.  The scale $\mu^2$ is taken to be 1 GeV${}^2$ and
serves only to define $c_0$.

Because these non-analytic contributions are complemented by terms
analytic in the quark mass or pion-mass squared, the slope and
curvature at large $m_\pi^2$ of these contributions is not
significant.  What is significant is the curvature at small $m_\pi^2$
and we see that this curvature is dominated by the LNA term.  Here
there is no mass splitting to mask the effects of dynamical chiral
symmetry breaking.  Thus, we will examine the extent to which our
simulation results are consistent with the LNA behaviour of \QchiPT.

The coefficient $\beta$ is related to the coefficient of the leading
non-analytic (LNA) contribution to the magnetic moment, $\chi$, via the
relation \cite{Leinweber:2002qb}
\begin{equation}
\beta \frac{m_N}{8\pi f_\pi^2} = \chi\ . 
\label{chibeta}
\end{equation} 
The coefficients $\chi$ have been determined for octet baryons and
their individual quark sector contributions in
Ref.~\cite{Leinweber:2002qb} and numerical values are reproduced in
Tables~\ref{tab:quarkMomChi} and \ref{tab:baryonMomChi} for ready
reference.

Since $m_\pi^2 <1\ {\rm GeV}^2$ in our simulations, the logarithmic
term is negative for all quark masses considered here.
Hence, the charge radius will exhibit a logarithmic divergence in
chiral limit to either positive or negative infinity, depending on the
whether $\beta$ (or $\chi$) is negative or positive respectively.

In the quenched approximation, the flavour-singlet $\eta'$ meson
remains degenerate with the pion and makes important contributions to
quenched chiral non-analytic behaviour.  The neutrality of its charge
prevents it from contributing to the coefficients of
Tables~\ref{tab:quarkMomChi} and \ref{tab:baryonMomChi}.  However, the
double hair-pin diagram in which the vector current couples to the
virtual baryon intermediary does give rise to chirally-singular
behaviour.  However the relatively small couplings render these
contributions small at the quark masses probed here.

\begin{table}[tbp] 
\caption{Coefficients, $\chi$, providing the LNA contribution to
baryon magnetic moments and charge radii in quenched QCD.
Coefficients for magnetic moments in full QCD are also indicated.
Here the coefficients for quark sector contributions to baryon
properties are indicated for quarks having unit charge.  Note that
$u_p$ for example denotes the coefficient for the two $u$ quarks of
the proton, each of which have unit charge.  Intermediate (Int.)
meson-baryon channels are indicated to allow for $SU(3)$-flavour
breaking in both the meson and baryon masses.  The coefficients are
calculated from the expressions of Ref.~\cite{Leinweber:2002qb} with
the axial couplings $F=0.50$ and $D=0.76$ with $f_\pi = 93$ MeV.}
\label{tab:quarkMomChi}
\addtolength{\tabcolsep}{-1pt}
\begin{ruledtabular}
\begin{tabular}{lccc}
$q$   &Int.           &Full QCD  &Quenched QCD \\
\hline 
\noalign{\smallskip}
$u_p \mid d_n$   &$N \pi$        &$- 6.87$   &$- 3.33$ \\     
                 &$\Lambda K$    &$- 3.68$   &$     0$ \\     
                 &$\Sigma  K$    &$- 0.15$   &$     0$ \\     
\noalign{\smallskip}                        
$d_p \mid u_n$   &$N \pi$        &$+ 6.87$   &$+ 3.33$ \\     
                 &$\Sigma  K$    &$- 0.29$   &$     0$ \\     
\noalign{\smallskip}                        
$s_p \mid s_n$   &$\Lambda K$    &$+ 3.68$   &$     0$ \\     
                 &$\Sigma K$     &$+ 0.44$   &$     0$ \\     
\noalign{\smallskip}
\hline 
\noalign{\smallskip}
$u_{\Sigma^+} \mid d_{\Sigma^-}$
                 &$\Sigma  \pi$  &$- 2.16$   &$     0$ \\     
                 &$\Lambda \pi$  &$- 1.67$   &$     0$ \\     
                 &$N K        $  &$     0$   &$- 0.29$ \\     
                 &$\Xi K      $  &$- 6.87$   &$- 3.04$ \\     
\noalign{\smallskip}                                   
$d_{\Sigma^+} \mid u_{\Sigma^-}$   
                 &$\Sigma  \pi$  &$+ 2.16$   &$     0$ \\     
                 &$\Lambda \pi$  &$+ 1.67$   &$     0$ \\     
                 &$N K        $  &$+ 0.29$   &$     0$ \\     
\noalign{\smallskip}
$s_{\Sigma}$     &$N K        $  &$- 0.29$   &$+ 0.29$ \\     
                 &$\Xi K      $  &$+ 6.87$   &$+ 3.04$ \\     
                 &$\Sigma \eta_{s}$  
                                 &$     0$   &$     0$ \\     
\noalign{\smallskip}
$u_{\Sigma^0} \mid d_{\Sigma^0}$   
                 &$\Sigma  \pi$  &$     0$   &$     0$ \\     
                 &$\Lambda \pi$  &$     0$   &$     0$ \\     
                 &$N K        $  &$+ 0.15$   &$- 0.15$ \\     
                 &$\Xi K      $  &$- 3.43$   &$- 1.52$ \\     
\noalign{\smallskip}
\hline 
\noalign{\smallskip}
$u_\Lambda \mid d_\Lambda$
                 &$\Sigma \pi $  &$     0$   &$     0$ \\    
                 &$\Lambda\eta_{l}$
                                 &$     0$   &$     0$ \\
                 &$N K        $  &$+ 3.68$   &$+ 1.23$ \\    
                 &$\Xi K      $  &$- 0.40$   &$+ 0.44$ \\    
\noalign{\smallskip}                                     
$s_\Lambda$      &$\Lambda\eta_{s}$
                                 &$     0$   &$     0$ \\    
                 &$N K        $  &$- 7.36$   &$- 2.45$ \\    
                 &$\Xi K      $  &$+ 0.79$   &$- 0.88$ \\    
\noalign{\smallskip}
\hline
\noalign{\smallskip}
$u_{\Xi^0} \mid d_{\Xi^-}$      
                 &$\Xi \pi$      &$- 0.29$   &$     0$ \\     
                 &$\Lambda K$    &$     0$   &$- 0.40$ \\     
                 &$\Sigma K$     &$+ 6.87$   &$+ 3.43$ \\     
                 &$\Omega K$     &$     0$   &$+ 0.29$ \\     
\noalign{\smallskip}                                
$d_{\Xi^0} \mid u_{\Xi^-}$      
                 &$\Xi \pi$      &$+ 0.29$   &$     0$ \\     
                 &$\Lambda K$    &$+ 0.40$   &$     0$ \\     
                 &$\Sigma K$     &$+ 3.43$   &$     0$ \\     
\noalign{\smallskip}                                
$s_{\Xi}$        &$\Lambda K$    &$- 0.40$   &$+ 0.40$ \\     
                 &$\Sigma K$     &$- 10.3$   &$- 3.43$ \\     
                 &$\Omega K$     &$     0$   &$- 0.29$ \\     
                 &$\Xi \eta_{s}$     
                                 &$     0$   &$     0$ \\     
\end{tabular}
\end{ruledtabular}
\end{table}

\begin{table}[p]
\caption{Coefficients, $\chi$, providing the LNA contribution to
baryon magnetic moments and charge radii in quenched QCD.
Coefficients for magnetic moments in full QCD are also indicated.
Intermediate (Int.) meson-baryon channels are indicated to allow for
$SU(3)$-flavour breaking in both the meson and baryon masses.  The
coefficients are calculated from the expressions of
Ref.~\cite{Leinweber:2002qb} with the axial couplings $F=0.50$ and
$D=0.76$ with $f_\pi = 93$ MeV.}
\label{tab:baryonMomChi}
\begin{ruledtabular}
\begin{tabular}{lccc}
Baryon     &Channel        &Full QCD   &Quenched QCD \\
\hline 
$p$        &$N \pi$        &$- 6.87$   &$- 3.33$ \\     
           &$\Lambda K$    &$- 3.68$   &$     0$ \\     
           &$\Sigma  K$    &$- 0.15$   &$     0$ \\     
\noalign{\bigskip}                               
$n$        &$N \pi$        &$+ 6.87$   &$+ 3.33$ \\     
           &$\Lambda K$    &$     0$   &$     0$ \\     
           &$\Sigma  K$    &$- 0.29$   &$     0$ \\     
\noalign{\bigskip}                               
$\Sigma^+$ &$\Sigma  \pi$  &$- 2.16$   &$     0$ \\     
           &$\Lambda \pi$  &$- 1.67$   &$     0$ \\     
           &$N K        $  &$     0$   &$- 0.29$ \\     
           &$\Xi K      $  &$- 6.87$   &$- 3.04$ \\     
           &$\Sigma \eta_{s}$
                           &$     0$   &$     0$ \\     
\noalign{\bigskip}                               
$\Sigma^0$ &$\Sigma  \pi$  &$     0$   &$     0$ \\     
           &$\Lambda \pi$  &$     0$   &$     0$ \\     
           &$N K        $  &$+ 0.15$   &$- 0.15$ \\     
           &$\Xi K      $  &$- 3.43$   &$- 1.52$ \\     
           &$\Sigma \eta_{s}$
                           &$     0$   &$     0$ \\     
\noalign{\bigskip}                               
$\Sigma^-$ &$\Sigma  \pi$  &$+ 2.16$   &$     0$ \\     
           &$\Lambda \pi$  &$+ 1.67$   &$     0$ \\     
           &$N K        $  &$+ 0.29$   &$     0$ \\     
           &$\Xi K      $  &$     0$   &$     0$ \\     
           &$\Sigma \eta_{s}$
                           &$     0$   &$     0$ \\     
\noalign{\bigskip}
$\Lambda$  &$\Sigma \pi $  &$     0$   &$     0$ \\    
           &$\Lambda\eta_{l}$
                           &$     0$   &$     0$ \\
           &$N K        $  &$+ 3.68$   &$+ 1.23$ \\    
           &$\Xi K      $  &$- 0.40$   &$+ 0.44$ \\    
           &$\Lambda\eta_{s}$
                           &$     0$   &$     0$ \\    
\noalign{\smallskip}
$\Xi^0$    &$\Xi \pi$      &$- 0.29$   &$     0$ \\     
           &$\Lambda K$    &$     0$   &$- 0.40$ \\     
           &$\Sigma K$     &$+ 6.87$   &$+ 3.43$ \\     
           &$\Omega K$     &$     0$   &$+ 0.29$ \\     
           &$\Xi    \eta_{s}$
                           &$     0$   &$     0$ \\    
\noalign{\bigskip}                               
$\Xi^-$    &$\Xi \pi$      &$+ 0.29$   &$     0$ \\     
           &$\Lambda K$    &$+ 0.40$   &$     0$ \\     
           &$\Sigma K$     &$+ 3.43$   &$     0$ \\     
           &$\Omega K$     &$     0$   &$     0$ \\     
           &$\Xi    \eta_{s}$
                           &$     0$   &$     0$ \\    
\end{tabular}
\end{ruledtabular}
\end{table}

\subsubsection{Quark sector charge radii}

We begin with an examination of the quark contributions to baryon
charge radii.  The results are reported for single quarks of unit
charge.  Of particular interest are the contributions of similar
quarks experiencing different environments.  Traditionally, quark
models of hadron structure neglected such environment sensitivity.
However, such environment sensitivity is manifest in chiral effective
field theory.  The finite kaon mass in the chiral limit renders the
kaon's contributions to curvature almost trivial relative to the pion.

\begin{figure}[tbp!]
\begin{center}
 {\includegraphics[height=\hsize,angle=90]{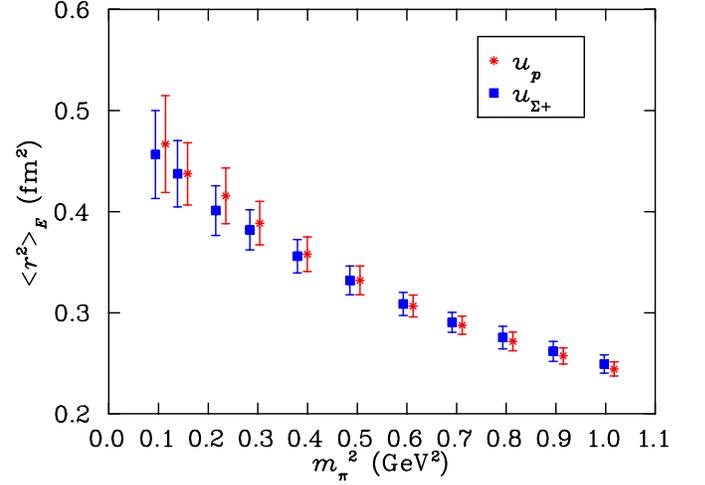}}
\end{center}
\vspace*{-0.5cm}
\caption{Electric charge radii of the $u$-quark distribution in the
  proton, ${u_p}$, and $\Sigma^+$, ${u}_{\Sigma^+}$, as a function of
  $m_\pi^2$ representing the quark masses considered in the
  simulation.  The results for ${u_p}$ are offset for clarity.}
\vspace{-0.2cm}
\label{crupus}
\end{figure}

Figure~\ref{crupus} displays the charge radii of the ${u}$-quark
distribution in the proton and compares this with the ${u}$-quark
distribution in $\Sigma^+$.  The $SU(3)$-flavour limit is manifest at
$m_\pi^2 \sim 0.5$ GeV${}^2$.  The replacement of a $d$-quark in the
proton, by an $s$ quark in $\Sigma^+$ gives rise to only a small
environment sensitivity in the $u$-quark properties.

Referring to the chiral coefficients of Table~\ref{tab:quarkMomChi},
the negative value of $\chi$ for $u_p$ indicates that the charge
radius of the $u$ quark distribution in the proton should diverge to
positive infinity in the chiral limit.
A physical understanding of this is made obvious by considering the
virtual transition $p \to n \pi^+$, which at the quark level can be
understood as $(uud) \to (udd) (\overline{d}u)$.
In the chiral limit, the $\pi^+$ carries the $u$ quark to infinity
such that $u$-quark charge distribution radius in the proton diverges.

In the case of the $u$ quark in $\Sigma^+$, the coefficient of the
logarithmic divergence is zero in the $\pi$ channel and hence no
divergence is expected.  While there is a significant coefficient for
transitions to $\Xi K$, the increased mass of the $\Xi$ baryon makes
this channel unfavourably suppressed.

The results in Fig.~\ref{crupus} for $u_p$ exhibit an upward trend and
increasing curvature with reducing quark mass.  The $u_{\Sigma^+}$
rises more slowly.  Hence these results are in qualitative agreement
with the LNA expectations of chiral effective field theory.

\begin{figure}[tbp!]
\begin{center}
 {\includegraphics[height=\hsize,angle=90]{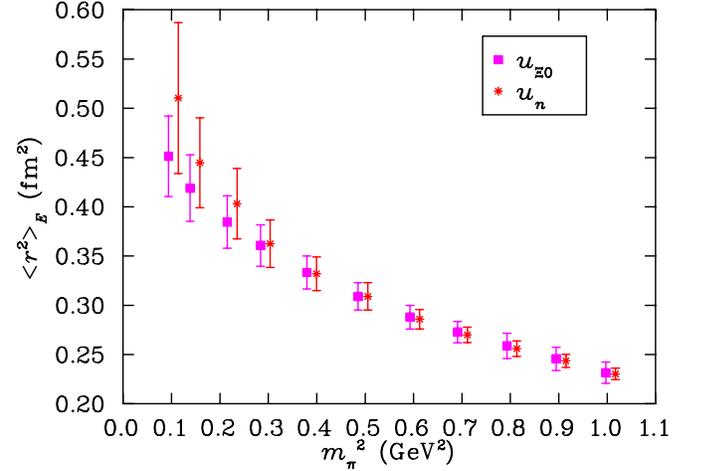}}
\end{center}
\vspace*{-0.5cm}
\caption{Charge radii of ${u_n}$ and ${u}_{\Xi^0}$ as a function of
  $m_\pi^2$.  The data for the ${u_n}$ are offset for clarity.}
\vspace{-0.2cm}
\label{crunux}
\end{figure}

Figure~\ref{crunux} displays the electric charge radii of the $u$
quark in the neutron ($u_n$) and in $\Xi^0$ ($u_{\Xi^0}$) as a
function of $m_{\pi}^2$.
Here we observe that the charge radii of $u_n$ and $u_{\Xi^0}$ are
nearly equal at heavy quark masses, but in the chiral limit they
differ.  The light $d$-quark environment of the ${u}$ quark in the
neutron provides enhanced chiral curvature as the chiral limit is
approached.

However, the true nature of the underlying physics is much more
subtle.  From Table~\ref{tab:quarkMomChi}, we see that the quenched
coefficient (last column) for the ${u}$ quark in the neutron is
positive in the $\pi$ channel, from which we can deduce that the
charge radius should actually diverge to negative infinity in the
chiral limit.

Physically this can be understood by looking at the quark
contributions to the virtual transition $n \to p \pi^-$ which gives
rise to this divergence.  In this case one has $(ddu) \to (duu) 
(\overline{u}d)$.
In the chiral limit, the mass of the pion approaches zero such that
the $\pi^-$ carries a $\overline{u}$ quark to infinity.
Since the $d$ quark is ignored while calculating the $u$ quark
contribution ({\it i.e.} the electric charge of the $d$ quark may be
thought of as zero), the entire charge of the pion comes from the
$\overline{u}$ quark, thus taking the $u$-quark charge distribution
radius to negative infinity.  However, Fig.~\ref{crunux} shows no such
trend.

The coefficient $\chi$ for $u_{\Xi^0}$ is zero in the $\pi$
channel, indicating that there should be no logarithmic divergence in
the chiral limit.
However it does have a substantial positive coefficient in the
favourable $\Sigma K$ channel, indicating the possibility of downward
curvature as the chiral limit is approached.  Again, Fig.~\ref{crunux}
shows no hint of downward curvature.

While the statistical error bars are sufficiently large to hide such a
turn over, there are other interpretations.  One possibility is that
we are not yet in the true chiral regime where such physics is
manifest.  Indeed, the divergence of the $u$-quark charge distribution
to negative infinity may only reveal itself at quark masses {\it
lighter} than the physical quark masses.

Alternatively, one might regard this particular case to be somewhat
exceptional.  It is the only channel in which chiral-loop physics is
expected to oppose the natural broadening of a distribution's Compton
wavelength.  On the lattice, the finite volume restricts the low
momenta of the effective field theory to discrete values.  It may be
that this lattice artifact prevents one from building up sufficient
strength in the loop integral to counter the Compton broadening.  In
this case it would be impossible to observe the divergence of $u_n \to
-\infty$ at any quark mass.  It will be interesting to resolve this
discrepancy with quantitative effective field theory calculations.

\begin{figure}[tbp!]
\begin{center}
 {\includegraphics[height=\hsize,angle=90]{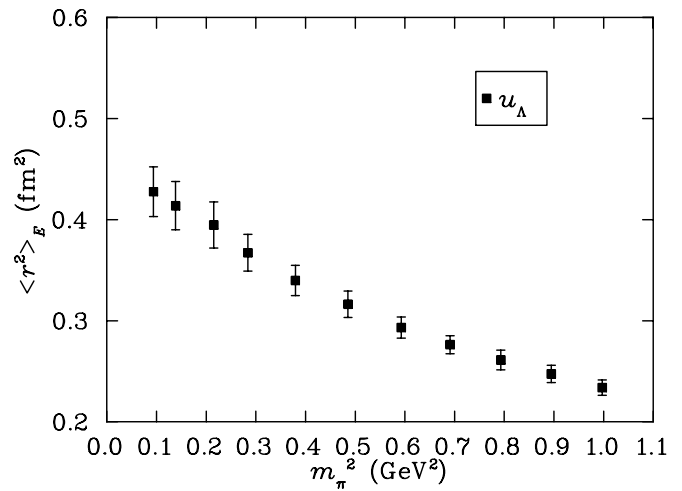}}
\end{center}
\vspace*{-0.5cm}
\caption{Electric charge distribution radius of a $u$ quark in
  $\Lambda$ as a function of $m_\pi^2$.}
\vspace{-0.2cm}
\label{crul}
\end{figure}

Figure~\ref{crul} reports our results for the charge distribution
radius of a $u$ quark in $\Lambda$ as a function of $m_\pi^2$.
The chiral coefficient for this is zero in the $\pi$ channel and hence
no divergence is expected.  However, there is significant strength for
downward curvature in the energetically favourable $NK$ channel.  
Indeed, the approach to the chiral limit is remarkably linear and
contrasts the upward curvature observed for other light quark
flavours.  Hence our results are in qualitative agreement with the
expectations of \QchiPT.

\begin{figure}[tbp!]
\begin{center}
 {\includegraphics[height=\hsize,angle=90]{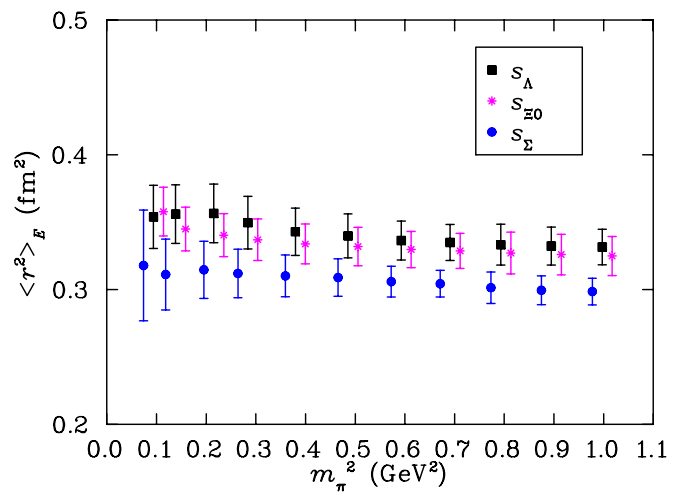}}
\end{center}
\vspace*{-0.5cm}
\caption{Electric charge distribution radii of strange quarks
  including ${s}_{\Lambda}$, ${s}_{\Xi^0}$ and ${s}_{\Sigma^0}$.  The
  data for ${s}_{\Xi^0}$ and ${s}_\Lambda$ are plotted at shifted
  ${m}^2_\pi$ values for clarity.}
\vspace{-0.2cm}
\label{crslxs}
\end{figure}

Figure~\ref{crslxs} illustrates the charge distribution
radius of strange quarks in $\Lambda^0$, $\Xi^0$ and $\Sigma^0$.  In
our simulations the strange quark mass is held fixed and therefore any
variation observed in the results is purely environmental
in origin.  All three distributions suggest a gentle dependence on the
mass of the environmental light quarks.

However, the environmental flavour-symmetry dependence of the strange
quark distributions is absolutely remarkable.  When the environmental
quarks are in an isospin 0 state in the $\Lambda$, the strange quark
distribution is broad.  On the other hand, when the environmental
quarks are in an isospin 1 state in $\Sigma$ baryons, the distribution
radius is significantly smaller.

In the case of strange quark distributions, the LNA contributions are
exclusively from transitions involving the kaon.  Therefore
significant curvature is not expected.  On the other hand, one might
expect broader distributions in cases where a virtual transition is
possible in quenched QCD.
Referring to Table~\ref{tab:quarkMomChi}, one sees that both
$s_\Lambda$ and $s_{\Xi}$ have strong transitions to the energetically
favourable $KN$ and $K\Sigma$ channels respectively.  The coefficients
are negative such that the virtual transitions will act to enhance the
charge distributions.  This is not the case for $s_\Sigma$ where the
sign is positive and the transition is to the energetically unfavoured
$K\Xi$ channel.  In summary, \QchiPT\ suggests the charge
distributions for $s_\Lambda$ and $s_\Xi$ will be larger than for
$s_\Sigma$.  This is exactly what is observed in Figure~\ref{crslxs}.

\subsubsection{Baryon charge radii}

\begin{figure}[tbp!]
\begin{center}
 {\includegraphics[height=\hsize,angle=90]{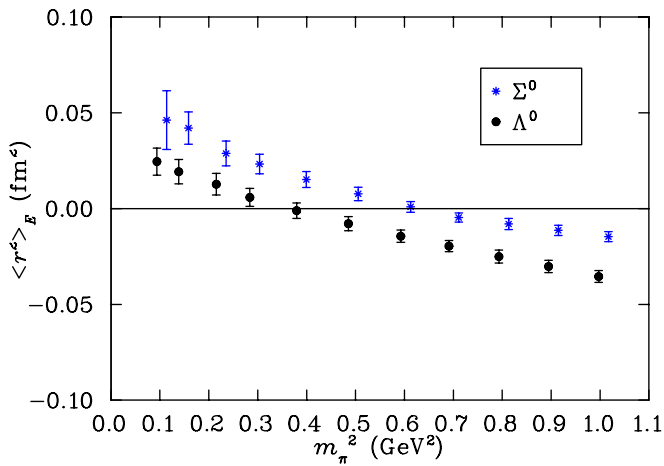}}
\end{center}
\vspace*{-0.5cm}
\caption{Electric charge radii of  $\Sigma^0$ and $\Lambda^0$ as a
  function of $m_\pi^2$.  The data for $\Sigma^0$ is offset to the
  right for clarity.}
\vspace{-0.2cm}
\label{crsl}
\end{figure}

The flavour-symmetry dependence of $uds$-quark distributions in
$\Sigma^0$ and $\Lambda^0$ is particularly manifest in
Fig.~\ref{crsl}.  Here the interplay between the light-quark sector
with effective charge $1/3$ and the strange sector with charge $-1/3$
is revealed.

At the $SU(3)$ flavour-symmetric point ($m_\pi^2 \sim 0.5\ {\rm
  GeV}^2$) where the strange and light quarks have the same mass and
the $\Lambda$ and $\Sigma$ are degenerate in mass, neither charge
radius is zero.  This very nicely reveals different charge
distributions for the quark sectors described in the previous
section. 

In constituent quark models, this flavour dependence would be
described in terms of spin-dependent forces.  In the $\Lambda^0$ where
a scalar diquark can form between the non-strange pair, the charge
radius is dominated by the broader strange-quark distribution at the
$SU(3)$-flavour symmetric point.  This is contrasted by the $\Sigma^0$
where scalar-diquark pairing would occur between strange and
non-strange quarks, acting to constrict the strange quark distribution
in $\Sigma$ as seen in Fig.~\ref{crslxs}.  In addition, hyperfine
repulsion in the non-strange quark sector leads to a broader
distribution for the light quark sector as indicated in
Tables~\ref{tab:eradS} and \ref{tab:eradL}.  
As compelling as this discussion is, this line of reasoning suggests
the decuplet baryon states should have broader quark distributions
\cite{Leinweber:1993nr} as scalar-diquark clusters do not dominate
there.  However, preliminary results from an analysis of decuplet
baryon structure on the same lattice configurations explored here
\cite{newDecuplet}, do not reveal broader quark distributions.  For
this reason, we consider our discussion of virtual transitions in the
context of effective field theory in the previous section to be a more
relevant description of the underlying physics.

Ultimately, as the chiral limit is approached, the light quark
distribution broadens and dominates the charge radii for both baryons.
However, the charge distribution of the $\Sigma^0$ is much broader and
reflects our discussion of the quark sector contributions.  In
particular, the LNA contributions of \QchiPT\ act to suppress the
distribution of $u_\Lambda$ and enhance $s_\Lambda$, whereas the LNA
contributions to $\Sigma^0$ are relatively suppressed either by having
small coefficients or having energetically unfavourable transitions in
the kaon channel.  This suppression of $u_\Lambda$ and enhancement
$s_\Lambda$ combines to give a strong net effect of suppressing the
charge radius of the $\Lambda^0$.

\begin{figure}[tbp!]
\begin{center}
 {\includegraphics[height=\hsize,angle=90]{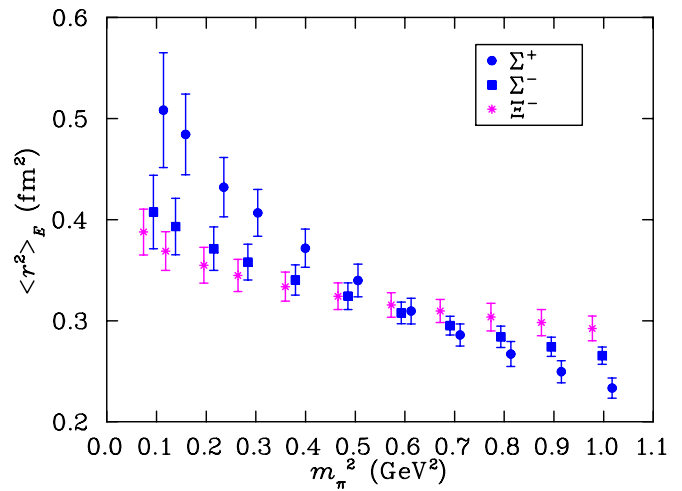}}
\end{center}
\vspace*{-0.5cm}
\caption{Electric charge radii of charged hyperons.  The $\Xi^-$ and
  $\Sigma^+$ are offset to the left and right respectively for
  clarity.}
\vspace{-0.2cm}
\label{crsx}
\end{figure}

The hyperon charge states, $\Sigma^-$ and $\Xi^-$ have chiral
coefficients which vanish in quenched QCD.  Similarly, $\Sigma^+$ has
no contributions from virtual pion transitions.  The one case, where
there is a substantial coefficient, is suppressed energetically.
Figure~\ref{crsx} displays our simulation results for the electric
charge distribution radii of these hyperons as a function of
$m_\pi^2$.  The ordering of the charge radii as the chiral limit is
approached is explained by the more localized strange quark
distribution.

\begin{figure}[tbp!]
\begin{center}
 {\includegraphics[height=\hsize,angle=90]{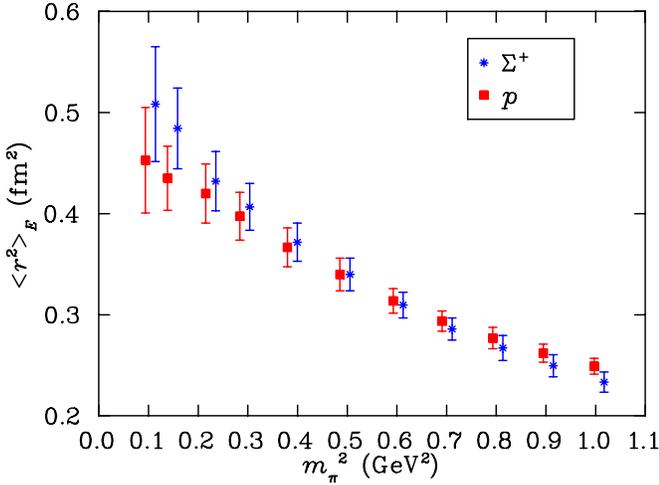}}
\end{center}
\vspace*{-0.5cm}
\caption{Electric charge radii of the proton and $\Sigma^+$.  Charge
  radii for $\Sigma^+$ are offset for clarity.}
\vspace{-0.2cm}
\label{crps}
\end{figure}

Figure~\ref{crps} compares the charge radii of $\Sigma^+$ with the
proton.  The charge radii of these baryons match at the SU(3) flavour
limit where $m_\pi^2 \sim 0.5\ {\rm GeV}^2$ as required.  As the
chiral limit is approached, the smaller charge distribution of the
heavier negatively-charged strange quark acts to make the $\Sigma^+$
larger.  This is manifest in the simulation results.

While the $\Sigma^+$ is not expected to display chiral curvature, the
proton charge radius presents one of the more favourable opportunities
to observe a hint of the logarithmic divergence to be encountered in
the chiral and infinite-volume limits of quenched QCD.
However, there is no hint of chiral curvature in favour of the proton
over the $\Sigma^+$.  

The origin of this discrepancy is once again traced to the
singly-represented $u$-quark in the neutron, or more specifically in
this case, the singly-represented $d$-quark in the proton.  As
highlighted in the discussion surrounding Fig.~\ref{crupus}, there is
a hint of increased curvature for the doubly-represented $u$-quark in
the proton over that in $\Sigma$, in accord with chiral effective
field theory.  But this is hidden in the proton charge radius due to
the absence of the anticipated curvature of the singly-represented
quark in the nucleon, as highlighted in the discussion surrounding
Fig.~\ref{crunux}.

\begin{figure}[tbp!]
\begin{center}
 {\includegraphics[height=\hsize,angle=90]{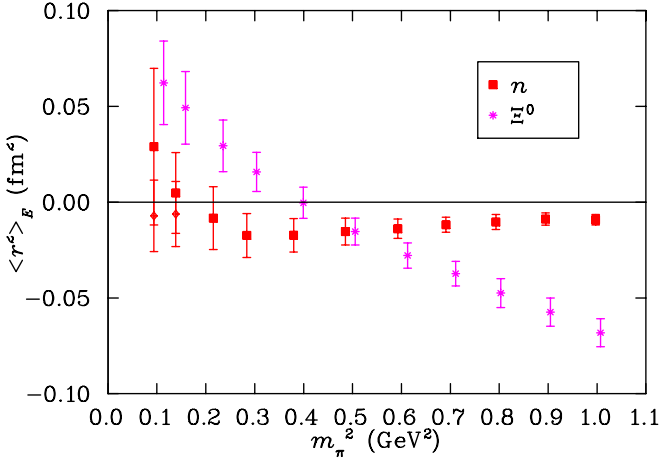}}
\end{center}
\vspace*{-0.5cm}
\caption{Electric charge radii of the neutron and $\Xi^0$. Charge
  radii for the $\Xi^0$ are shifted to the right for clarity.
  Asymmetric error bars for the neutron charge radius are described in
  the text.}
\vspace{-0.2cm}
\label{crnx}
\end{figure}

Similarly, the ultimate divergence of the neutron charge radius to
negative infinity via $n \to p \pi^-$ is not yet manifest.  Rather a
crossing of the central values into positive values of squared charge
radii is revealed in Fig.~\ref{crnx}.  Still, the statistical errors
remain consistent with negative values.

The crossing of the central values of the squared neutron charge
radius into positive values led us to further examine our selection of
fit regime in our correlation function analysis.  Our concern is that
noise in the correlation function may be distorting the fit.  Hence,
we have also considered fits including $t=15$, immediately following
the point-split current insertion centred at $t=14$.  While we prefer
to allow some Euclidean time evolution following the current
insertion, this systematic uncertainty is reflected in the asymmetric
error bar of Fig.~\ref{crnx} for the lightest two neutron charge
radii.

\begin{figure}[tbp!]
\begin{center}
 {\includegraphics[height=\hsize,angle=90]{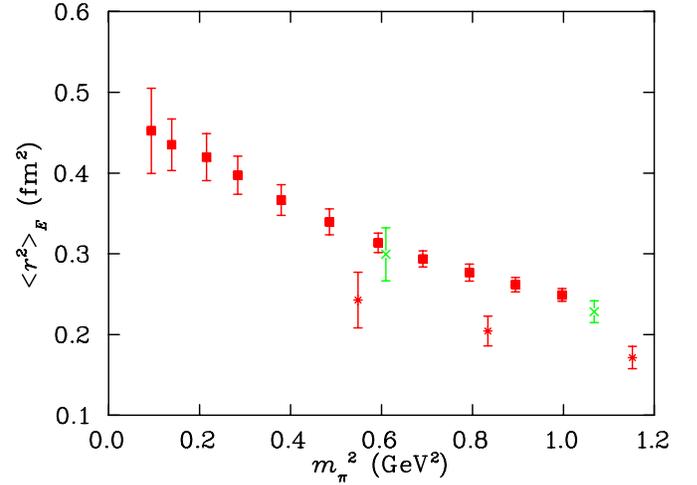}}
\end{center}
\vspace*{-0.5cm}
\caption{The proton charge radius is compared with previous state of
  the art lattice simulation results in quenched QCD.  The solid
  squares indicate current lattice QCD results with FLIC fermions.
  The stars indicate the lattice results of \cite{Leinweber:1990dv}
  while the crosses indicate the results of \cite{Wilcox:1991cq}, both
  of which use the standard Wilson actions for the gauge and fermion
  fields.}  
\vspace{-0.5cm}
\label{crp}
\end{figure}

To summarise, we have explored the electric form factors of the baryon
octet and their quark sector contributions at light quark masses
approaching the chiral regime.  The unprecedented nature of our quark
masses is illustrated in Fig.~\ref{crp} which compares the present
results for the proton charge radius with the previous state of the
art \cite{Leinweber:1990dv,Wilcox:1991cq}.  Here the static quark
potential has been used to uniformly set the scale among all the
results.  The small values of the early results are most likely due to
the small physical lattice volumes necessitated at that time.  
The precision afforded by 400 $20^3 \times 40$ lattices is manifest. 

We have discovered that all baryons having non-vanishing
energetically-favourable couplings to virtual meson-baryon transitions
tend to be broader than those which do not.  This qualitative
realisation provides a simple explanation for the patterns revealed in
our quenched-QCD simulations.

Still, evidence of chiral {\it curvature} on our large-volume lattice
is rather subtle in general and absent in the exceptional case of the
singly-represented quark in the neutron or $\Xi$.  In this case, it is
thought that the restriction of momenta to discrete values on the
finite-volume lattice prevents the build up of strength in the loop
integral of effective field theory.  Without sufficient strength, the
Compton broadening of the distribution will not be countered as the
chiral limit is approached.

\subsection{Magnetic moments}

The magnetic moment $\mu$ is provided by the magnetic form factor at
$Q^2 = 0$, ${\cal{G}}_M(0)$, with units of the natural magneton,
$\mu_B = {e}/({2\, M_B})$, where $M_B$ is the mass of the baryon
\begin{equation}
\mu = {\cal{G}}_M(0)\, \frac{e}{2M_B} \, .
\end{equation} 
\label{magmom}
While we could present a detailed discussion of the magnetic form
factors summarised in Sec.~\ref{ffCorr}, a more interesting discussion
of the results is facilitated via the magnetic moment where chiral
non-analytic behaviour takes on a simple functional form and a vast
collection of phenomenology is available to provide a context for our
results. 

Since the magnetic form factors must be calculated at a finite value
of momentum transfer, $Q^2$, the magnetic moment must be inferred from
our results, ${\cal{G}}_M(Q^2)$, obtained at the minimum non-vanishing
momentum transfer available on our periodic lattice.  The $Q^2$
dependence of lattice results from the QCDSF collaboration
\cite{Gockeler:2003ay} are described well by a dipole.
Phenomenologically this is a well established fact for the nucleon at
low momentum transfers.  

However, we will take an even weaker approximation and assume only
that the $Q^2$ dependence of the electric and magnetic form factors is
similar, without stating an explicit functional form for the $Q^2$
dependence.  This too is supported by experiment where the proton
ratio $\frac{{\cal{G}}_M(Q^2)}{\mu \, {\cal{G}}_E(Q^2)} \simeq 1$ for
values of $Q^2$ similar to that probed here.  In this case
\begin{equation}
{\cal{G}}_M(0) =
\frac{{\cal{G}}_M(Q^2)}{{\cal{G}}_E(Q^2)} \, {\cal{G}}_E(0) \, .
\label{pffm}
\end{equation}
The strange and light sectors of hyperons will scale differently, and
therefore we apply Eq.~(\ref{pffm}) to the individual quark sectors.
Octet baryon properties are then reconstructed as described in the
discussion surrounding Eq.~(\ref{reconsEg}) in Sec.~\ref{ffCorr}.
Results for baryon magnetic moments and their quark sector
contributions are summarised in Tables~\ref{tab:magmomN} through
\ref{tab:magmomX}.

\begin{table*}[tbp] 
\caption{Nucleon magnetic moments and their quark sector
  contributions.  Sector contributions are indicated for single
  quarks having unit charge.  Baryon charge states are also
  summarised.  Values for $m_\pi^2$ and magnetic moments are in
  units of ${\rm GeV}^2$ and $\mu_N$ respectively.}
\label{tab:magmomN}
\begin{ruledtabular}
\begin{tabular}{ccccc}
$m_\pi^2$    &  $u_p$       & $d_p$          & $p$          & $n$ \\
\hline
$0.9972(55)$ & $0.960(13)$  &  $-0.366(8)$   & $1.401(18)$  & $-0.883(11)$ \\
$0.8947(54)$ & $0.995(15)$  &  $-0.373(9)$   & $1.451(21)$  & $-0.913(12)$ \\
$0.7931(53)$ & $1.032(18)$  &  $-0.382(10)$  & $1.503(25)$  & $-0.943(14)$ \\
$0.6910(35)$ & $1.064(14)$  &  $-0.386(10)$  & $1.547(19)$  & $-0.967(12)$ \\
$0.5925(33)$ & $1.108(17)$  &  $-0.395(12)$  & $1.610(24)$  & $-1.002(15)$ \\ 
$0.4854(31)$ & $1.163(24)$  &  $-0.406(16)$  & $1.686(33)$  & $-1.046(20)$ \\
$0.3795(31)$ & $1.231(29)$  &  $-0.424(20)$  & $1.783(41)$  & $-1.104(25)$ \\
$0.2839(33)$ & $1.317(41)$  &  $-0.433(28)$  & $1.901(57)$  & $-1.167(36)$ \\
$0.2153(35)$ & $1.395(62)$  &  $-0.447(47)$  & $2.009(87)$  & $-1.228(56)$ \\
$0.1384(43)$ & $1.517(79)$  &  $-0.456(60)$  & $2.17(11)$   & $-1.315(71)$ \\
$0.0939(44)$ & $1.54(11)$  &  $-0.521(87)$  & $2.22(15)$   & $-1.372(92)$ 
\end{tabular}
\end{ruledtabular}

\caption{$\Sigma$ magnetic moments and their quark sector
  contributions.  Sector contributions are indicated for single
  quarks having unit charge.  Baryon charge states are also
  summarised.  Values for $m_\pi^2$ and magnetic moments are in units
  of ${\rm GeV}^2$ and $\mu_N$ respectively.}
\label{tab:magmomS}
\begin{ruledtabular}
\begin{tabular}{cccccc}
$m_\pi^2$    & $u_{\Sigma^+}$ &  $s_\Sigma$    & $\Sigma^+$   & $\Sigma^0$    & $\Sigma^-$ \\
\hline
$0.9972(55)$ & $0.976(15)$  &  $-0.416(13)$  & $1.440(21)$  & $0.464(7)$    & $-0.512(10)$ \\
$0.8947(54)$ & $1.010(18)$  &  $-0.414(14)$  & $1.484(24)$  & $0.474(8)$    & $-0.535(12)$ \\ 
$0.7931(53)$ & $1.044(21)$  &  $-0.412(15)$  & $1.530(28)$  & $0.485(9)$    & $-0.559(14)$ \\
$0.6910(35)$ & $1.072(15)$  &  $-0.408(12)$  & $1.565(21)$  & $0.493(7)$    & $-0.579(11)$ \\
$0.5925(33)$ & $1.113(18)$  &  $-0.407(14)$  & $1.620(26)$  & $0.507(8)$    & $-0.607(13)$ \\ 
$0.4854(31)$ & $1.163(24)$  &  $-0.406(16)$  & $1.686(33)$  & $0.523(10)$   & $-0.640(16)$ \\
$0.3795(31)$ & $1.221(27)$  &  $-0.409(18)$  & $1.764(38)$  & $0.543(12)$   & $-0.678(18)$ \\
$0.2839(33)$ & $1.286(34)$  &  $-0.412(21)$  & $1.852(47)$  & $0.566(14)$   & $-0.720(22)$ \\
$0.2153(35)$ & $1.344(42)$  &  $-0.421(27)$  & $1.932(58)$  & $0.588(18)$   & $-0.756(28)$ \\
$0.1384(43)$ & $1.418(50)$  &  $-0.421(34)$  & $2.031(72)$  & $0.613(23)$   & $-0.805(32)$ \\
$0.0939(44)$ & $1.446(77)$  &  $-0.429(42)$  & $2.07(11)$   & $0.625(30)$   & $-0.821(53)$ 
\end{tabular}
\end{ruledtabular}
\end{table*}

\begin{table*}[tbp] 
\caption{The $\Lambda^0$ magnetic moment and its quark sector
  contributions.  Sector contributions are indicated for single
  quarks having unit charge.  Baryon charge states are also
  summarised.  Values for $m_\pi^2$ and magnetic moments are in units
  of ${\rm GeV}^2$ and $\mu_N$ respectively.}
\label{tab:magmomL}
\begin{ruledtabular}
\begin{tabular}{cccc}
$m_\pi^2$    & $u_\Lambda$  & $s_\Lambda$     & $\Lambda^0 $\\
\hline
$0.9972(55)$ & $0.086(10)$  &  $1.611(29)$    & $-0.509(9)$  \\
$0.8947(54)$ & $0.091(12)$  &  $1.621(20)$    & $-0.510(10)$ \\
$0.7931(53)$ & $0.095(13)$  &  $1.631(32)$    & $-0.512(10)$ \\
$0.6910(35)$ & $0.102(11)$  &  $1.650(21)$    & $-0.516(8)$  \\
$0.5925(33)$ & $0.109(12)$  &  $1.666(24)$    & $-0.519(8)$  \\ 
$0.4854(31)$ & $0.117(14)$  &  $1.688(28)$    & $-0.524(10)$ \\
$0.3795(31)$ & $0.124(16)$  &  $1.715(32)$    & $-0.530(11)$ \\
$0.2839(33)$ & $0.135(18)$  &  $1.749(37)$    & $-0.538(13)$ \\
$0.2153(35)$ & $0.140(24)$  &  $1.780(43)$    & $-0.547(16)$ \\
$0.1384(43)$ & $0.151(27)$  &  $1.799(49)$    & $-0.549(18)$ \\
$0.0939(44)$ & $0.154(31)$  &  $1.804(53)$    & $-0.550(20)$ 
\end{tabular}
\end{ruledtabular}

\caption{$\Xi$ magnetic moments and their quark sector
  contributions.  Sector contributions are indicated for single
  quarks having unit charge.  Baryon charge states are also
  summarised.  Values for $m_\pi^2$ and magnetic moments are in units
  of ${\rm GeV}^2$ and $\mu_N$ respectively.}
\label{tab:magmomX}
\begin{ruledtabular}
\begin{tabular}{ccccc}
$m_\pi^2$    & $s_\Xi$       &  $u_{\Xi^0}$   & $\Xi^0$        & $\Xi^-$ \\
\hline
$0.9972(55)$ & $1.132(32)$   &  $-0.361(13)$  & $-0.996(23)$   & $-0.635(22)$ \\
$0.8947(54)$ & $1.137(33)$   &  $-0.371(14)$  & $-1.005(24)$   & $-0.634(23)$ \\
$0.7931(53)$ & $1.141(34)$   &  $-0.381(15)$  & $-1.015(24)$   & $-0.634(24)$ \\
$0.6910(35)$ & $1.152(21)$   &  $-0.385(13)$  & $-1.025(18)$   & $-0.640(14)$ \\
$0.5925(33)$ & $1.157(22)$   &  $-0.395(14)$  & $-1.035(19)$   & $-0.640(15)$ \\ 
$0.4854(31)$ & $1.163(24)$   &  $-0.406(16)$  & $-1.046(20)$   & $-0.640(16)$ \\
$0.3795(31)$ & $1.172(24)$   &  $-0.421(18)$  & $-1.062(21)$   & $-0.641(17)$ \\
$0.2839(33)$ & $1.181(25)$   &  $-0.437(20)$  & $-1.079(23)$   & $-0.642(17)$ \\
$0.2153(35)$ & $1.189(27)$   &  $-0.454(23)$  & $-1.096(25)$   & $-0.642(18)$ \\
$0.1384(43)$ & $1.199(29)$   &  $-0.475(27)$  & $-1.116(29)$   & $-0.641(20)$ \\
$0.0939(44)$ & $1.212(29)$   &  $-0.495(31)$  & $-1.138(32)$   & $-0.643(20)$ 
\end{tabular}
\end{ruledtabular}
\end{table*}

\subsubsection{Quenched chiral perturbation theory}

As for the charge radii, it is interesting to compare our results with
the LNA and NLNA terms of \chiPT\ which survive to some extent in
\QchiPT.  As for the charge radii, the NLNA contributions provide
little curvature \cite{Young:2004tb} and we turn our attention to the
LNA contributions \cite{Leinweber:2002qb}.  These LNA contributions to
baryon magnetic moments have their origin in couplings of the
electromagnetic current to the virtual meson propagating in the
intermediate meson-baryon state.

For virtual pion transitions, the LNA terms have the very simple form
$\chi\, m_\pi \sim {m_q}^{1/2}$, with values for $\chi$ as summarised
in Tables~\ref{tab:quarkMomChi} and \ref{tab:baryonMomChi}.  While
this contribution is finite in the chiral limit, the rate of change of
this contribution does indeed diverge in the chiral limit.  The less
singular nature of this contribution should allow its contributions to
be observed at larger pion masses, making magnetic moments an
excellent observable to consider in searching for evidence of chiral
curvature.  Kaon contributions take on the same form in the limit in
which baryon mass splittings are neglected.

As for the charge radii, negative values of $\chi$ provide curvature
towards more positive values as the chiral limit is approached, and
vice versa for positive values of $\chi$.

As emphasised earlier in our discussion of charge radii, the
flavour-singlet $\eta'$ meson remains degenerate with the pion in the
quenched approximation and makes important contributions to quenched
chiral non-analytic behaviour.  The neutrality of its charge prevents it
from contributing to the coefficients of Tables~\ref{tab:quarkMomChi}
and \ref{tab:baryonMomChi}.  However, the double hair-pin diagram in
which the vector current couples to the virtual baryon intermediary
does give rise to a logarithmic divergence in baryon magnetic moments.
However the relatively small couplings of the $\eta'$ render these
contributions negligible at the quark masses probed here
\cite{Young:2004tb}.

\subsubsection{Quark sector magnetic moments}

%
\begin{figure}[tbp!]
\begin{center}
 {\includegraphics[height=\hsize,angle=90]{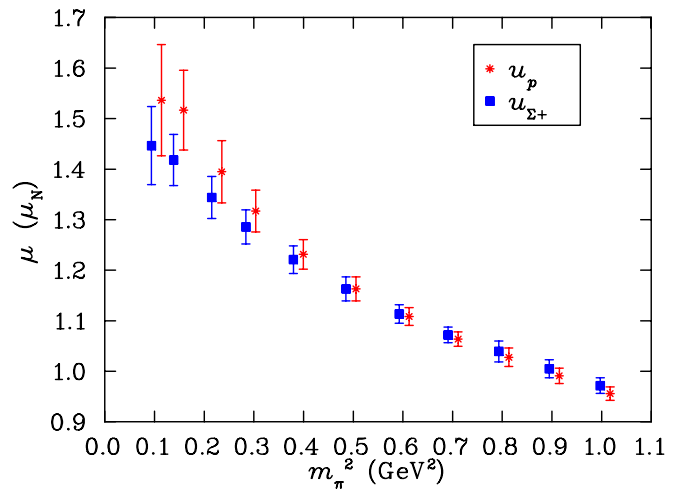}}
\end{center}
\vspace*{-0.5cm}
\caption{Magnetic moment contributions of the $u$-quark sector to the
  proton, ${u_p}$, and the $\Sigma^+$, ${u}_{\Sigma^+}$.  The
  contributions ${u_p}$ are shifted right for clarity.}
\vspace{-0.5cm}
\label{magmomups}
\end{figure}

The $u$-quark contribution to the proton and $\Sigma^+$ magnetic
moments are illustrated in Fig.~\ref{magmomups}.  The contribution
$u_p$ was described as the most optimal channel for directly observing
chiral non-analytic curvature in quenched lattice QCD simulations
\cite{Leinweber:2002qb} and this curvature is evident in
Fig.~\ref{magmomups}.

The value of $\chi$ for $u_p$ is large and negative, predicting LNA
curvature towards positive values as the chiral limit is approached.
The value of $\chi$ for $u_{\Sigma^+}$ vanishes in the $\pi$ channel.
Similarly, strength in the $\Xi K$ channel is energetically
suppressed.  Hence the chiral curvature is predicted to be negligible
for $u_{\Sigma^+}$ and will contrast the upward curvature for $u_p$.
This is observed in our lattice simulations.  Figure ~\ref{magmomups}
reveals curvature in $u_p$ relative to a rather linear approach for
$u_{\Sigma^+}$ to the chiral limit.

\begin{figure}[tbp!]
\begin{center}
 {\includegraphics[height=\hsize,angle=90]{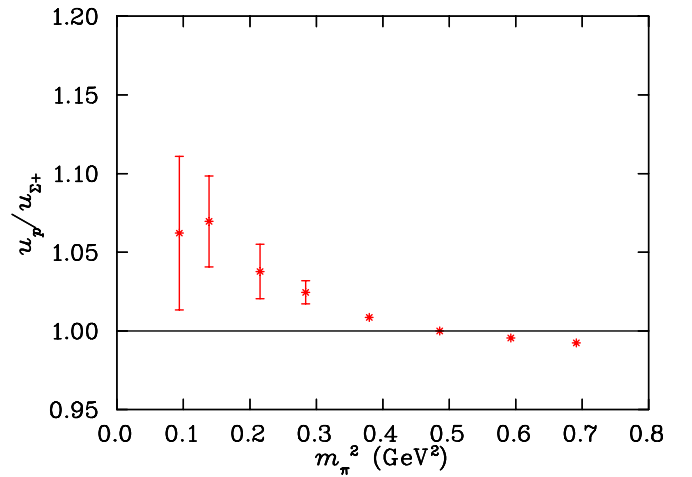}}
\end{center}
\vspace*{-0.5cm}
\caption{Ratio of $u$-quark contributions to the magnetic moments of
   the proton and $\Sigma^+$.  Chiral curvature in the $u$-quark
  contribution to the proton's moment gives rise
  to significant enhancement in $u_p$.}
\vspace{-0.5cm}
\label{magmomUpOnUSigma}
\end{figure}

The results for $u_p$ and $u_{\Sigma^+}$ are highly correlated and
therefore the enhancement of the magnetic moment of $u$ in the proton
over the $\Sigma^+$ provides significant evidence of chiral
non-analytic behaviour in accord with the LNA predictions of chiral
perturbation theory.  The strong correlation of these results is
evident in the $SU(3)$ flavour-symmetric point at $m_\pi^2 \simeq 0.5\
{\rm GeV}^2$ where the results are identical.  To expose the
significance of this result, we present Fig.~\ref{magmomUpOnUSigma}
illustrating the correlated ratio of magnetic moment contributions
$u_p/u_{\Sigma^+}$.  There, the significance exceeds two standard
deviations for quark masses between the lightest quark mass considered
and the $SU(3)$ flavour limit at $m_\pi^2 \simeq 0.5\ {\rm GeV}^2$.

%
\begin{figure}[tbp!]
\begin{center}
 {\includegraphics[height=\hsize,angle=90]{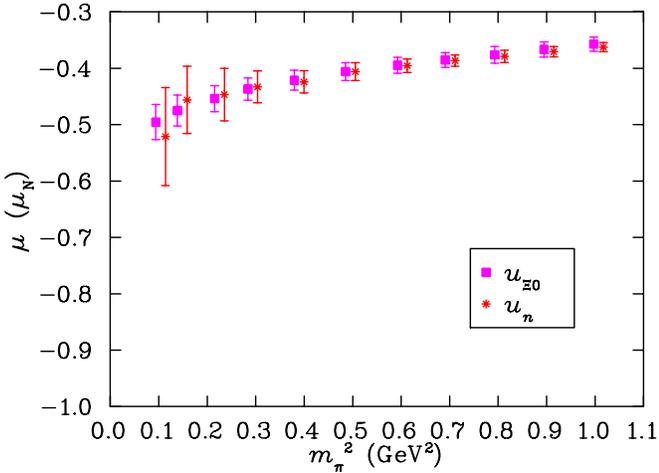}}
\end{center}
\vspace*{-0.5cm}
\caption{The u-quark contribution (single quark of unit charge) to the
  magnetic moments of the neutron, ${u_n}$, and $\Xi^0$,
  ${u}_{\Xi^0}$.  The magnetic moment for ${u_n}$ is shifted to the
  right for clarity.}
\vspace{-0.5cm}
\label{magmomunx}
\end{figure}

Figure~\ref{magmomunx} illustrates the magnetic moment contribution of
the single $u$ quark in the neutron and the $\Xi^0$, normalised to
unit charge.  The magnetic moments match at the $SU(3)$-flavour limit
where $m_\pi^2 \simeq 0.5\ {\rm GeV}^2$ as required.  The environment
sensitivity of the $u$ quark contribution is subtle and is most
evident in the size of the statistical error bar.

The chiral coefficient, $\chi$, of the non-analytic term $\sim m_\pi$
for $u_n$ is large and greater than zero, predicting curvature towards
negative values as the chiral limit is approached.  While the
coefficient for $u_{\Xi^0}$ vanishes in the $\pi$ channel, a
substantial coefficient resides in the energetically favoured $\Sigma
K$ channel and therefore some curvature towards negative values are
again predicted as the chiral limit is approached.
Fig.~\ref{magmomunx} is in accord with these predictions for the
chiral curvature.

We note that for this case of magnetic moments, the anticipated chiral
curvature is indeed observed for this sector.  This contrasts the case
of charge distribution radii, where chiral curvature was to oppose the
Compton-broadening of the distribution and was not manifest in the
simulation results.

\begin{figure}[tbp!]
\begin{center}
 {\includegraphics[height=\hsize,angle=90]{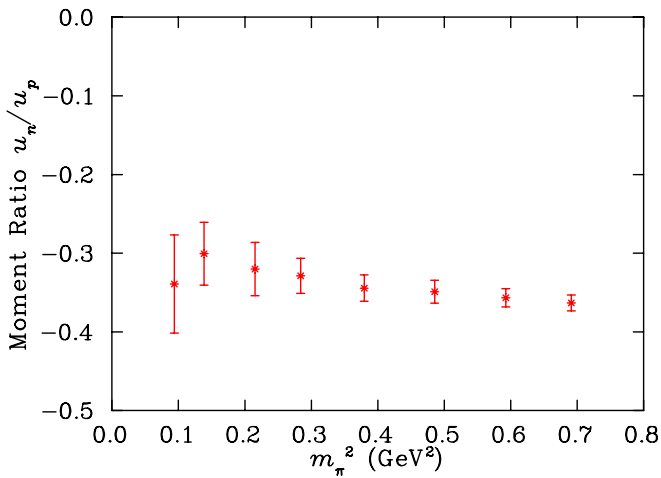}}
\end{center}
\vspace*{-0.5cm}
\caption{The ratio of singly ($u_n$) and doubly ($u_p$) represented
  quark contributions for single quarks of unit charge to nucleon
  magnetic moments.  In the simple $SU(6)$ spin-flavour symmetric
  quark model the predicted ratio is constant at $-1/2$.}
\label{unupRatio}
\end{figure}

It is interesting to examine the ratio of of singly ($u_n$) and doubly
($u_p$) represented quark contributions (for single quarks of unit
charge) to nucleon magnetic moments \cite{Leinweber:1990dv}.  The
$SU(6)$ spin-flavour symmetry of the simple quark model provides
\begin{eqnarray}
\mu_p &=& \frac{4}{3}\, \mu_u - \frac{1}{3}\, \mu_d \,  \\
&=& \frac{2}{3} \, 2\, \frac{2}{3}\, \mu_q^{\rm QM}
  - \frac{1}{3} \, 1\, \left ( -\frac{1}{3} \right) \, \mu_q^{\rm QM} \, 
\label{prefac}
\end{eqnarray}
where $\mu_q^{\rm QM}$ is the constituent quark moment.  The quark
moment pre-factors in Eq.~(\ref{prefac}) are respectively, $SU(6)$,
quark number and charge factors.  Discarding quark number and charge
factors, one arrives at the SU(6) prediction for $u_n/u_p$ for single
quarks of unit charge of $-1/2$.  This prediction is to be compared
with Fig.~\ref{unupRatio} which reveals this ratio to be substantially
smaller than the $SU(6)$ prediction, even at the $SU(3)$
flavour-symmetric limit where $m_\pi^2 \simeq 0.5\ {\rm GeV}^2$.
This result is in accord with Ref.~\cite{Leinweber:1990dv} where this
effect was first observed in lattice QCD.

The gentle slope of the results in Fig.~\ref{unupRatio} at larger
quark masses suggests that the $SU(6)$ spin-flavour symmetric quark
model prediction of $-1/2$ will be realised only at much heavier quark
masses than those examined here.

\begin{figure}[tbp!]
\begin{center}
 {\includegraphics[height=\hsize,angle=90]{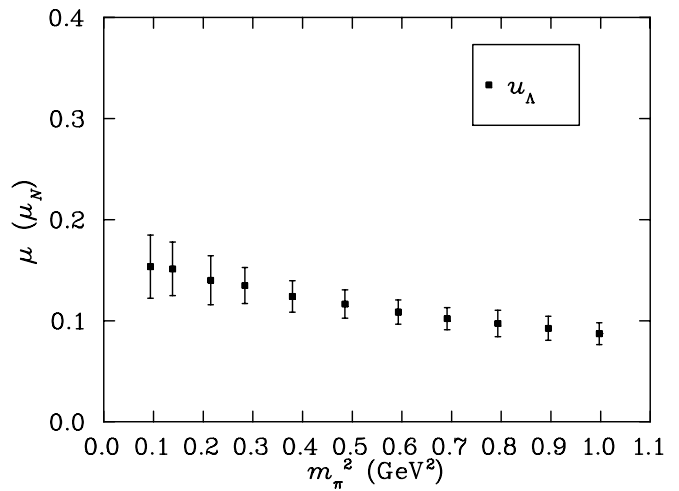}}
\end{center}
\vspace*{-0.5cm}
\caption{Magnetic moment contribution of the $u$-quark sector (or
equivalently the $d$-quark sector) to the $\Lambda^0$ magnetic
moment.}
\label{magmomul}
\end{figure}

%
Figure~\ref{magmomul} shows the magnetic moment contribution of the
$u$-quark sector (or equivalently the $d$-quark sector) to the
$\Lambda^0$ magnetic moment, normalised for a single quark of unit
charge.  In simple quark models, this contribution is zero as the $u$
and $d$ quarks are in a spin-0, isospin-0 state.  Our simulation
results reveal that the dynamics of QCD, even in the quenched
approximation, are much more complex.  The contribution of $u_\Lambda$
differs from zero by more than eight standard deviations at the $SU(3)$
flavour-symmetric point, and confirms earlier findings
\cite{Leinweber:1990dv} of a non-trivial role for the light quark
sector in the magnetic moment of $\Lambda^0$.

The chiral coefficient for $u_\Lambda$ vanishes in the pion channel
and has only small strength in the energetically favoured $NK$
channel.  Hence little curvature is anticipated and this is supported
by our findings in Fig.~\ref{magmomul}.


\begin{figure}[tbp!]
\begin{center}
 {\includegraphics[height=\hsize,angle=90]{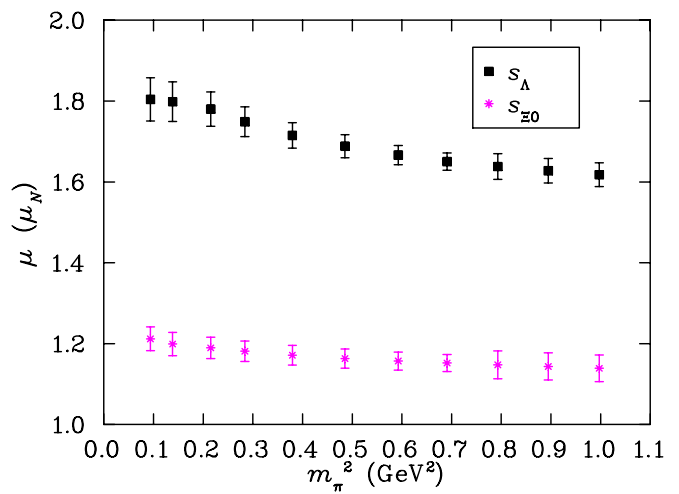}}
\end{center}
\vspace*{-0.5cm}
\caption{Magnetic moments of ${s}_{\Lambda}$ and ${s}_{\Xi^0}$ as a
  function of quark mass.}
\label{magmomslx}
\end{figure}

\begin{figure}[tbp!]
\begin{center}
 {\includegraphics[height=\hsize,angle=90]{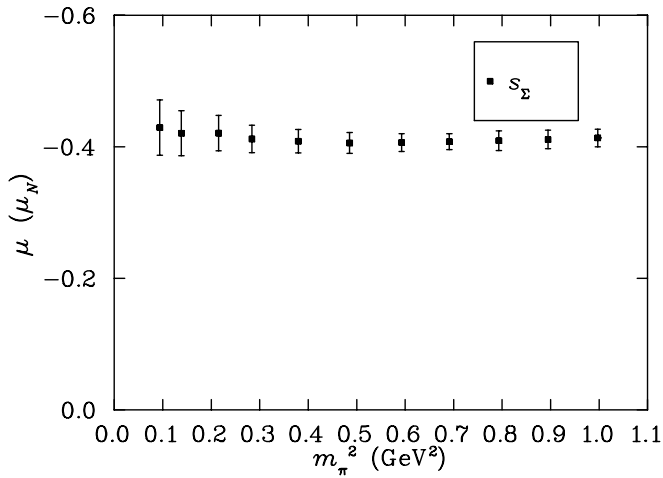}}
\end{center}
\vspace*{-0.5cm}
\caption{Magnetic moment contribution of the strange quark in
  $\Sigma$, ${s}_{\Sigma}$, as a function of quark mass.  }
\label{magmomss}
\end{figure}

Turning our attention to the strange quark sectors,
Figs.~\ref{magmomslx} and \ref{magmomss} present results for
$s_\Lambda$, $s_{\Xi}$ and $s_{\Sigma}$ magnetic moments.  In our
simulations the strange quark mass is held fixed and therefore any
variation observed in the results is purely environmental in origin.
While $s_{\Xi}$ and $s_{\Sigma}$ display only a mild environment
sensitivity, $s_\Lambda$ shows a remarkably-robust dependence on its
light quark environment.

Recall that in our examination of the environmental flavour-symmetry
dependence of the strange quark distribution, a strong sensitivity
was found.  When the environmental quarks are in an isospin-0 state in
the $\Lambda$, the strange quark distribution is broad.  On the other
hand, when the environmental quarks are in an isospin-1 state in
$\Sigma$ baryons, the distribution radius is significantly smaller.
It appears that the broad distribution of the strange quark in
$\Lambda$ makes it sensitive to the dynamics of its neighbours.

In the case of strange-quark moments, the LNA contributions are
exclusively from transitions involving the kaon.  Referring to
Table~\ref{tab:quarkMomChi}, one sees that both $s_\Lambda$ and
$s_{\Xi}$ have strong transitions to the energetically favourable $KN$
and $K\Sigma$ channels respectively.  The coefficients are negative
such that the virtual transitions will act to provide curvature
towards positive values, enhancing the magnetic moments in these
cases.  This is not the case for $s_\Sigma$ where the sign is positive
and the transition is to the energetically unfavoured $K\Xi$ channel.
In summary, \QchiPT\ suggests the magnetic moments for $s_\Lambda$ and
$s_\Xi$ will display curvature that acts to enhance the magnetic
moment whereas $s_\Sigma$ will display little curvature.  These
predictions are exactly as observed in Figs.~\ref{magmomslx} and
\ref{magmomss}.


\subsubsection{Baryon magnetic moments}

\begin{figure}[tbp!]
\begin{center}
 {\includegraphics[height=\hsize,angle=90]{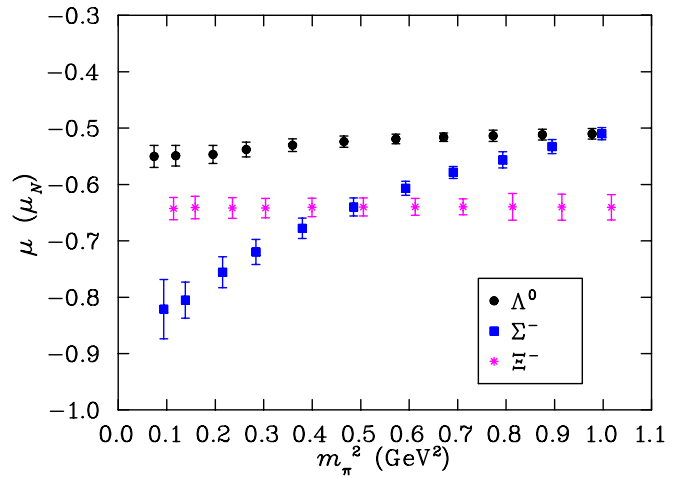}}
\end{center}
\vspace*{-0.5cm}
\caption{Magnetic moments of $\Sigma^-$, $\Lambda^0$ and $\Xi^-$.
  Results for $\Lambda^0$ and $\Xi^-$ are offset left and right
  respectively for clarity.}
\label{magmombslx}
\end{figure}

Figure~\ref{magmombslx} depicts the magnetic moments of $\Lambda^0$,
$\Sigma^{-}$ and $\Xi^{-}$.  As the magnetic moments of $\Lambda^0$
and $\Xi^{-}$ are dominated by the strange quark contribution, these
moments show only a gentle dependence on the quark mass.  These
contrast $\Sigma^-$ where the light $d$ quarks dominate the moment.

The hyperon charge states, $\Sigma^-$ and $\Xi^-$, have LNA chiral
coefficients which vanish in quenched QCD.  On the other hand, the
$\Lambda^0$ magnetic moment has some positive strength in the
energetically favoured $NK$ channel, suggesting curvature towards
negative values as the chiral limit is approached.  These features are
manifest in the $\Lambda^0$ and $\Xi^-$ moments of
Fig.~\ref{magmombslx} where the curvature in the $\Lambda^0$ moment
towards negative values contrasts the invariance of the $\Xi^-$
moment.  The approach of the $\Sigma^-$ moment to the chiral limit is
also fairly linear, in accord with a vanishing LNA chiral
coefficient.

\begin{figure}[tbp!]
\begin{center}
 {\includegraphics[height=\hsize,angle=90]{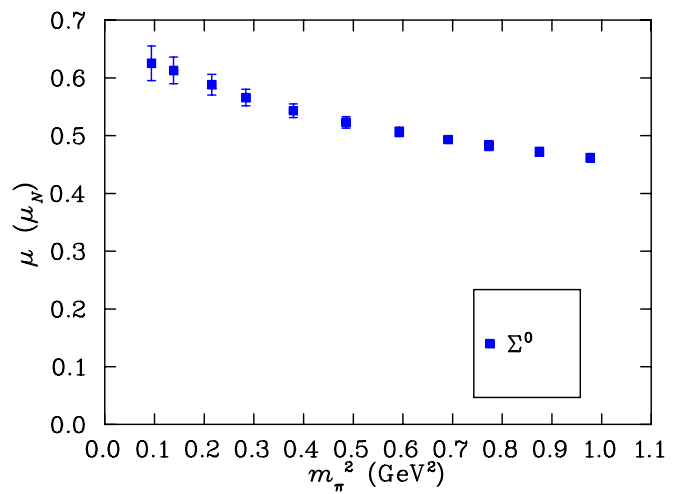}}
\end{center}
\vspace*{-0.5cm}
\caption{Magnetic moment of $\Sigma^0$.}
\label{magmombs}
\end{figure}

\begin{figure}[tbp!]
\begin{center}
 {\includegraphics[height=\hsize,angle=90]{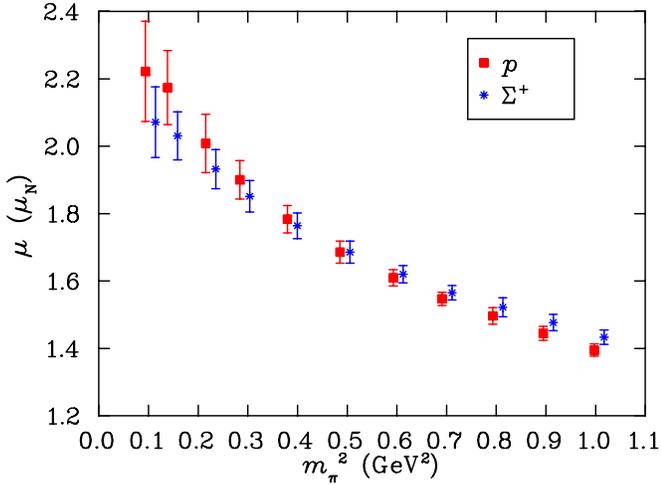}}
\end{center}
\vspace*{-0.5cm}
\caption{Magnetic moments of the proton and $\Sigma^+$.  The
  $\Sigma^+$ moments are offset to the right for clarity.}
\label{magmombps}
\end{figure}

Figure~\ref{magmombs} presents results for the $\Sigma^0$ baryon where
chiral curvature is anticipated to be small.  However, a comparison of
$p$ and $\Sigma^+$ magnetic moments provides a favourable opportunity
to observe chiral curvature.  The proton has a strong negative
coupling to the pion channel, predicting curvature towards positive
values as the chiral limit is approached.  This contrasts the
$\Sigma^+$ where the strong coupling is to the energetically
unfavourable $\Xi K$ channel suggesting a more linear approach to the
chiral limit.

Figure~\ref{magmombps} depicts the variation of these moments with
quark mass.  These results are highly correlated and therefore the
enhancement of the magnetic moment of the proton over the $\Sigma^+$
provides significant evidence of chiral non-analytic behaviour in accord
with the LNA predictions of chiral perturbation theory.  The strong
correlation of these results is evident in the $SU(3)$
flavour-symmetric point at $m_\pi^2 \simeq 0.5\ {\rm GeV}^2$ where the
results are identical.

\begin{figure}[tbp!]
\begin{center}
 {\includegraphics[height=\hsize,angle=90]{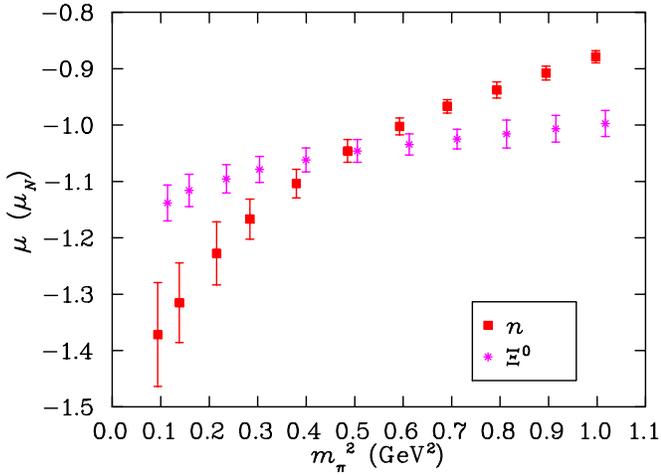}}
\end{center}
\vspace*{-0.5cm}
\caption{Magnetic moments of the neutron and $\Xi^0$.  $\Xi^0$ moments
  are offset to the right for clarity.}
\vspace{-0.5cm}
\label{magmombnx}
\end{figure}

Figure~\ref{magmombnx} reports the magnetic moments of the neutron and
$\Xi^0$.  The neutron provides a favourable case for the observation of
chiral curvature associated with the pion channel.  Similarly the
$\Xi^0$ has significant strength in the energetically favoured $\Sigma
K$ channel.  In both cases the chiral coefficient, $\chi$ is positive,
predicting curvature towards negative values as the chiral limit is
approached.  These predictions are in accord with the observations of
Fig.~\ref{magmombnx}.

\begin{figure}[tbp!]
\begin{center}
 {\includegraphics[height=\hsize,angle=90]{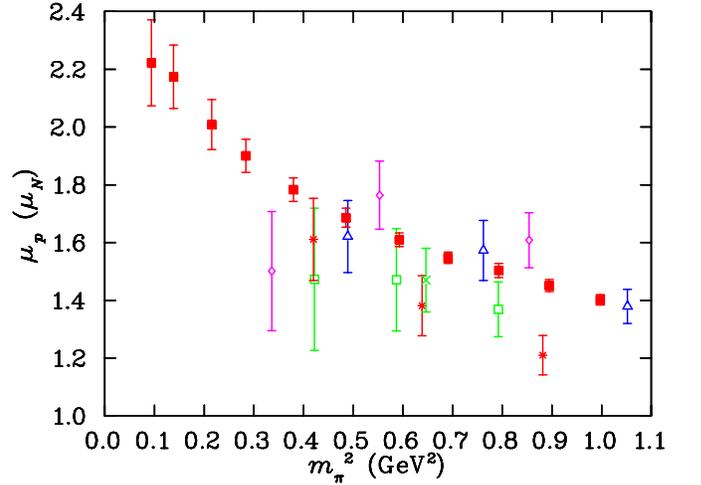}}
\end{center}
\vspace*{-0.5cm}
\caption{Proton magnetic moment in nuclear magnetons is compared from
  a variety of lattice simulations.  The solid squares indicate our
  current lattice QCD results with FLIC fermions.  The stars indicate
  the early lattice results of Ref.~\cite{Leinweber:1990dv}.  The
  crosses (only one point) indicate the results of
  Ref.~\cite{Wilcox:1991cq}.  The open symbols describe the QCDSF
  collaboration results \cite{Gockeler:2003ay}.  Open squares indicate
  results with $\beta=6.0$, open triangles indicate those with
  $\beta=6.2$ while the open diamonds indicate their results with
  $\beta=6.4$.}
\label{magmomprotcomp}
\end{figure}

In summary, we have performed an unprecedented exploration of the
light quark-mass properties of octet-baryon magnetic moments in
quenched QCD.  Figure~\ref{magmomprotcomp} presents our results in the
context of recent state of the art results from lattice QCD
\cite{Leinweber:1990dv,Wilcox:1991cq,Gockeler:2003ay}.  The precision
afforded by 400, $20^3 \times 40$ lattices and the efficient access to
the chiral regime enabled by our use of the FLIC fermion action is
clear.  In every case, the LNA curvature predicted by chiral
perturbation theory is manifest in our results.

\subsubsection{Ratio of $\mu_{\Xi^-}$ to $\mu_{\Lambda}$}

The experimentally measured value of magnetic moment of $\Xi^-$ is
$-0.651 \pm 0.003 \mu_N$ and that of $\Lambda$ is $-0.613 \pm 0.004
\mu_N$, making their ratio greater than 1 at $1.062 \pm 0.012$.  This
has presented a long-standing problem to constituent quark models
which predict the magnetic moment ratio, $\Xi^-/\Lambda^0$, to be less
than one.

In the simple $SU(6)$ spin-flavour quark model, the magnetic moment of
$\Xi^-$ is 
\begin{equation}
\mu_{\Xi^-} = \frac{4}{3}\mu_s - \frac{1}{3}\mu_d \, ,
\label{magmomXi}
\end{equation}
where $\mu_s$ and $\mu_d$ are the magnetic moments of the constituent
$s$ and $d$ quarks respectively.  Since the $u$-$d$ pair in $\Lambda$
forms a spin-$0$ state, the magnetic moment of the $\Lambda$ has a sole
contribution from the $s$ quark
\begin{equation}
\mu_{\Lambda} = \mu_s \, .
\label{magmoml}
\end{equation}
Taking the ratio yields
\begin{equation}
\frac{\mu_{\Xi^-}}{\mu_{\Lambda}} =  \frac{4}{3} - \frac{1}{3}
\frac{\mu_d}{\mu_s} \, .
\label{ratiomagmom}
\end{equation}
Now since, the magnetic moment of a charged Dirac particle goes
inversely as its mass, and since the $s$ and $d$ quarks have identical
charge, the ratio may be written
\begin{equation}
\frac{\mu_{\Xi^-}}{\mu_{\Lambda}} = \frac{4}{3} - \frac{1}{3}
\frac{m_s}{m_d} \, , 
\label{magmomCQrat}
\end{equation}
where $m_d$ and $m_s$ are the constituent masses of the $d$ and $s$
quarks respectively.  Given that $m_s > m_d$ it is inescapable that
this ratio is less than 1 in the simple quark model.  Indeed, the
accepted values of $d$ and $s$ constituent quark masses place this
ratio at $0.836$.

\begin{figure}[tbp!] 
\begin{center}
 {\includegraphics[height=\hsize,angle=90]{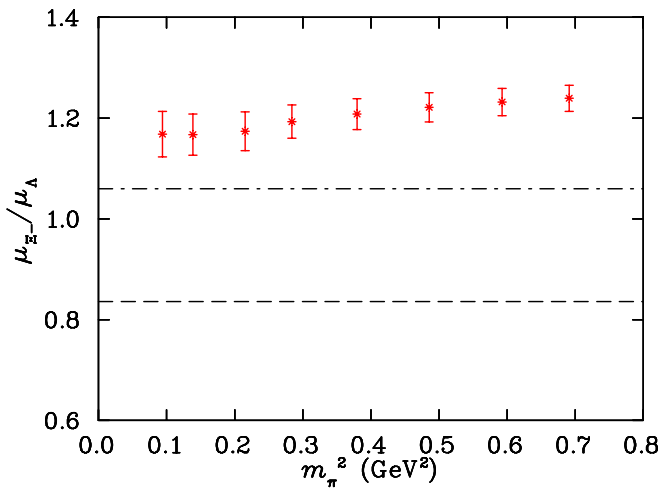}}
\end{center}
\vspace*{-0.5cm}
\caption{Magnetic moment ratio $\mu_{\Xi^-}/\mu_\Lambda$ as calculated
  in quenched QCD.  The simple quark model results of 0.836 is
  illustrated by the dashed line while the experimental value of the
  ratio is indicated by the dot-dash line.}
\vspace{-0.5cm}
\label{ratiomom}
\end{figure}

Figure~\ref{ratiomom} shows the ${\mu_{\Xi^-}}/{\mu_\Lambda}$
ratio as a function of quark mass as observed in our quenched lattice
calculations.  Remarkably, the ratio is greater than one at all quark
masses.

There are two important aspects of our previous discussion that give
rise to a result exceeding 1.  First, as illustrated in
Fig.~\ref{unupRatio}, we have found that the singly represented quarks
give a contribution to the magnetic moment that is much smaller in
magnitude than that of the $SU(6)$ quark model prediction.  This gives
rise to a 40\% reduction in the contribution of the second term of
Eq.~(\ref{magmomCQrat}).  

While this is sufficient to correct the ratio to $\sim 1$, there is a
second effect.  Namely, the light quark sector makes a non-trivial
contribution to the $\Lambda$ magnetic moment.  As illustrated in
Fig.~\ref{magmomul}, this contribution is positive for unit charge
quarks.  Since the net charge of the $u$-$d$ sector is $+1/3$, the
contribution of the $s$ quark in $\Lambda^0$ must have a negative
value whose magnitude exceeds the observed $\Lambda$ moment.  And this
is seen in Fig.~\ref{magmomslx}.  There, chiral curvature in
$s_\Lambda$ makes the ratio of magnetic moment contributions
$s_\Lambda/s_\Xi \sim 3/2$ as opposed to the $SU(6)$ suggestion of
4/3.  This resolves the long-standing discrepancy.

\subsection{Magnetic radii}

Using the values for the magnetic moments obtained by scaling the
individual quark sector contributions to $Q^2 = 0$, and our values for
the form factors at finite $Q^2$, magnetic radii may be determined in
exactly the same fashion as the electric radii.

\begin{figure}[tbp!]
\begin{center}
 {\includegraphics[height=\hsize,angle=90]{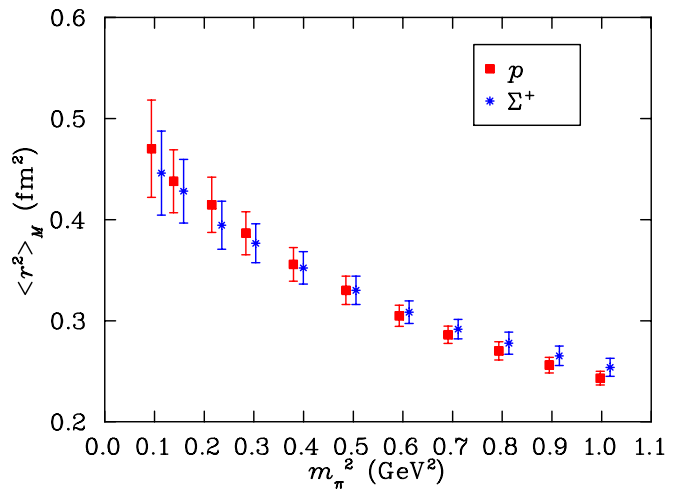}}
\end{center}
\vspace*{-0.5cm}
\caption{Magnetic radii of the proton and $\Sigma^+$.  The latter are
  shifted right for clarity.}
\label{magradps}
\end{figure}

\begin{figure}[tbp!]
\begin{center}
 {\includegraphics[height=\hsize,angle=90]{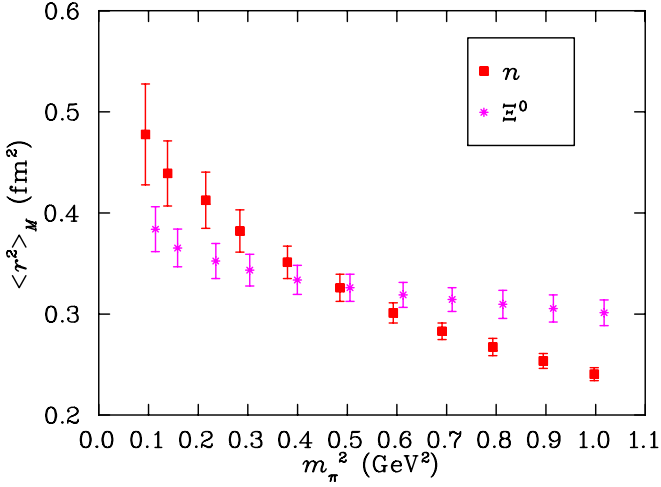}}
\end{center}
\vspace*{-0.5cm}
\caption{Magnetic radii of the neutron and $\Xi^0$.  The latter are
  shifted to the right for clarity.}
\label{magradnx}
\end{figure}

\begin{figure}[tbp!]
\begin{center}
 {\includegraphics[height=\hsize,angle=90]{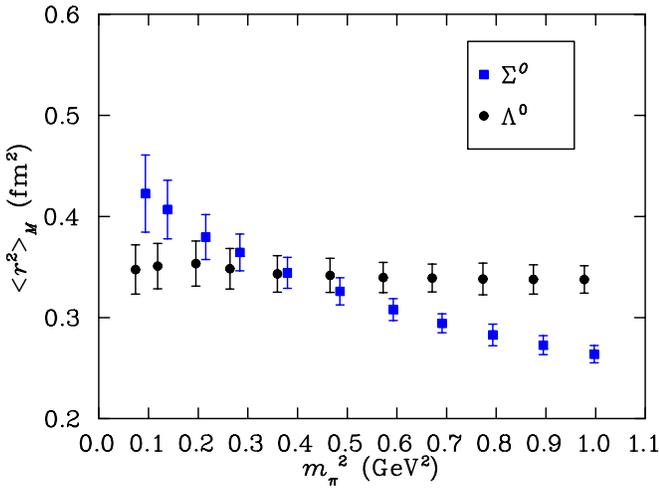}}
\end{center}
\vspace*{-0.5cm}
\caption{Magnetic radii of $\Sigma^0$ and $\Lambda$.  The latter are
  shifted left for clarity.}
\label{magradsl}
\end{figure}

\begin{figure}[tbp!]
\begin{center}
 {\includegraphics[height=\hsize,angle=90]{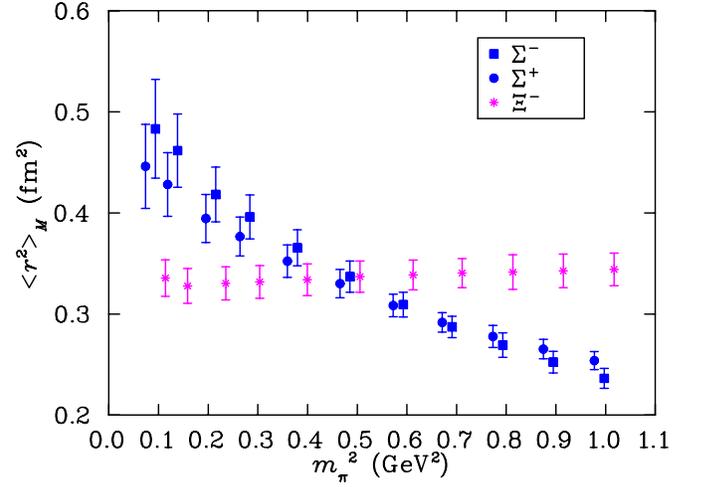}}
\end{center}
\vspace*{-0.5cm}
\caption{Magnetic radii of $\Sigma^+$, $\Sigma^-$ and $\Xi^-$.
  Results for $\Sigma^+$ and $\Xi^-$ are offset left and right
  respectively for clarity.  }
\label{magradsx}
\end{figure}

Analogous to the charge radius, we adopt a dipole form for the $Q^2$
dependence and define the magnetic radius as
\begin{equation}
\frac{\langle r_M^2 \rangle}{{\cal G}_M(0)} = {12 \over Q^2} \left (
    \sqrt{ {\cal G}_M(0) \over {\cal G}_M(Q^2) } - 1 \right ) .
\quad
\label{magrad}
\end{equation}
The magnetic radii, ${\langle r_M^2 \rangle}/{{\cal G}_M(0)}$, are
tabulated in Table~\ref{tab:magradB}.
Figures~\ref{magradps} through \ref{magradsx} display the variation of
the magnetic radii with $m_{\pi}^2$ for the octet baryons.

\begin{table*}[tbp]
\caption{Magnetic radii ${\langle r_M^2 \rangle}/{{\cal G}_M(0)}$ of
  the octet baryons in ${\rm fm}^2$ for various $m_\pi^2$ in ${\rm
  GeV}^2$.}
\label{tab:magradB}
\begin{ruledtabular}
\begin{tabular}{ccccccccc}
$m_\pi^2$    & $p$          & $n$          & $\Sigma^+$   & $\Sigma^0$   & $\Sigma^-$   & $\Lambda$    & $\Xi^0$      & $\Xi^-$ \\ 
\hline
$0.9972(55)$ & $0.241(7)$   & $0.240(6)$   & $0.254(9)$   & $0.264(9)$   & $0.236(10)$  & $0.328(14)$  & $0.298(13)$  & $0.337(16)$ \\
$0.8947(54)$ & $0.255(8)$   & $0.252(7)$   & $0.265(10)$  & $0.273(9)$   & $0.252(10)$  & $0.328(14)$  & $0.302(13)$  & $0.335(17)$ \\
$0.7931(53)$ & $0.269(9)$   & $0.266(9)$   & $0.278(10)$  & $0.283(10)$  & $0.269(12)$  & $0.328(16)$  & $0.306(13)$  & $0.334(17)$ \\
$0.6910(35)$ & $0.286(9)$   & $0.283(8)$   & $0.292(10)$  & $0.294(9)$   & $0.287(11)$  & $0.339(14)$  & $0.314(12)$  & $0.340(14)$ \\
$0.5925(33)$ & $0.305(10)$  & $0.301(10)$  & $0.308(11)$  & $0.308(11)$  & $0.309(12)$  & $0.340(15)$  & $0.319(12)$  & $0.339(15)$ \\
$0.4854(31)$ & $0.330(14)$  & $0.326(13)$  & $0.330(14)$  & $0.326(13)$  & $0.337(15)$  & $0.342(17)$  & $0.326(13)$  & $0.337(15)$ \\
$0.3795(31)$ & $0.356(17)$  & $0.351(16)$  & $0.352(16)$  & $0.344(15)$  & $0.365(18)$  & $0.343(18)$  & $0.334(14)$  & $0.334(16)$ \\
$0.2839(33)$ & $0.387(21)$  & $0.382(21)$  & $0.377(19)$  & $0.364(18)$  & $0.396(22)$  & $0.348(20)$  & $0.343(16)$  & $0.332(16)$ \\
$0.2153(35)$ & $0.415(27)$  & $0.413(28)$  & $0.395(24)$  & $0.380(22)$  & $0.418(27)$  & $0.353(22)$  & $0.352(17)$  & $0.330(16)$ \\
$0.1384(43)$ & $0.438(31)$  & $0.439(32)$  & $0.428(32)$  & $0.407(29)$  & $0.462(36)$  & $0.351(22)$  & $0.365(19)$  & $0.328(17)$ \\
$0.0939(44)$ & $0.470(48)$  & $0.478(50)$  & $0.446(42)$  & $0.423(38)$  & $0.483(49)$  & $0.347(24)$  & $0.384(22)$  & $0.336(18)$
\end{tabular}
\end{ruledtabular}
\end{table*}

Figure~\ref{magradps} depicts the magnetic radii of the proton and
$\Sigma^+$ as a function of input quark mass.  The somewhat subtle
differences have a simple explanation in terms of the more localised
strange quark in $\Sigma$.
 
In the proton, the long-range nature of the light-quark distributions
means that their contributions to the magnetic form factor reduce
quickly for increasing momentum transfers.  In the case of $\Sigma^+$,
which has a broadly distributed $u$ quark distribution and a narrowly
distributed $s$ quark distribution, the reduction in magnitude of the
form factor is less.  Here, the $s$-quark distribution contributes
positively and remains relatively invariant with increased resolution.
Thus the $\Sigma^+$ has a larger form factor than the proton at finite
$Q^2$ and hence a smaller magnetic radius.

Figure~\ref{magradnx} reports the magnetic radii of the neutron and
$\Xi^0$.  Following a similar argument as above, the neutron is
expected to have a larger magnetic radius than the $\Xi^0$, and this
is confirmed in the plot.

Figure~\ref{magradsl} illustrates the magnetic radii of $\Lambda$ and
$\Sigma^0$ as a function of the input quark mass.  In $\Lambda$, most
of the magnetic moment has its origin in the $s$ quark and therefore
the magnetic radius will be relatively small.  In the $\Sigma^0$ the
$u$-$d$ sector is a major contributor to the form factor.  As a result,
the form factor reduces more at finite momentum transfer, which in
turn implies that the magnetic radius of the $\Sigma^0$ will be
relatively large.

Figure~\ref{magradsx} illustrates the magnetic radii of $\Sigma^-$,
$\Sigma^+$ and $\Xi^-$.  $\Sigma^+$ is replotted here to facilitate
comparison with the other two members of the baryon octet.

$\Sigma^-$ has the largest magnetic radius among the octet baryons and
this is to be expected based on our considerations of the origin of
the baryon magnetic moment.  Here, the doubly represented $d$ quark
contributes to the total baryon form factor with the same sign,
whereas the strange sector acts to reduce the magnitude of the total
form factor.  Upon increasing the momentum transfer resolution, the
$d$-sector is reduced dramatically whereas the strange sector, acting
to reduce the total form factor, is relatively preserved.  this gives
rise to a large drop in the total form factor at finite $Q^2$ and thus
a large magnetic radius.

On the other hand, the singly represented $d$ quark in the $\Xi^-$
makes only a small contribution to the $\Xi^-$ form factor, and
therefore the magnetic radius reflects the small distribution of the
strange quark.

\section{SUMMARY}
\label{sec:Summary}

We have presented an extensive investigation of the electromagnetic
properties of octet baryons in quenched QCD.  The development of the
${\cal O}(a)$-improved FLIC fermion action has been central to
enabling this study.  The FLIC fermion operator is an efficient
nearest-neighbour fermion operator with excellent scaling properties
\cite{Zanotti:2004dr}.  The vastly improved chiral properties of this
operator \cite{FLIClqm} enables the exploration of the electromagnetic
form factors at quark masses significantly lighter than those
investigated in the past.  The unprecedented nature of our quark
masses is illustrated in Figs.~\ref{crp} and \ref{magmomprotcomp} for
the proton charge radii and magnetic moments, respectively.

Central to our discussion of the results is the search for evidence of
chiral non-analytic behaviour as predicted by chiral perturbation
theory.  We have discovered that all baryons having non-vanishing,
energetically-favourable couplings to virtual meson-baryon transitions
tend to be broader than those which do not.  This qualitative
realisation provides a simple explanation for the patterns revealed in
our quenched-QCD simulations.

Of particular interest is the environmental isospin dependence of the
strange quark distributions in $\Lambda^0$ and $\Sigma^0$.  When the
environmental quarks are in an isospin-0 state in the $\Lambda$, the
strange quark distribution is broad.  On the other hand, when the
environmental quarks are in an isospin-1 state in $\Sigma$ baryons,
the distribution radius is significantly smaller.

Still, evidence of chiral {\it curvature} on our large-volume lattice
is rather subtle in general and absent in the exceptional case of the
singly-represented quark in the neutron or $\Xi$.  In this case, the
chiral loop effects act to oppose the Compton broadening of the
distribution.  However, it is thought that the restriction of momenta
to discrete values on the finite-volume lattice prevents the build up
of strength in the loop integral sufficient to counter the natural
broadening of the distribution as the quark becomes light.  It will be
interesting to explore this quantitatively in finite-volume chiral
effective field theory.

In contrast, chiral curvature is evident in the quark-sector
contributions to baryon magnetic moments.  In every case, the
curvature predicted by chiral perturbation theory is manifest in our
results.  Of particular mention is the comparison of the $u$-quark
contribution to the proton and $\Sigma^+$ illustrated in
Figs.~\ref{magmomups} and \ref{magmomUpOnUSigma}.  The environment
sensitivity of the $s$ quark in $\Lambda^0$ depicted in
Fig.~\ref{magmomslx} is particularly robust.

We find it remarkable that the leading non-analytic features of chiral
perturbation theory are observed in our simulation results.  Naively,
one might have expected a non-trivial role for the higher order terms
of the chiral expansion which might have acted to hide the leading
behaviour.  However, the smooth and slow variation of our simulation
results indicate that these higher order terms must sum to provide
only a small correction to the leading behaviour.  These observations
indicate that regularisations of chiral effective field theory which
resum the chiral expansion at each order, to ensure that higher order
terms sum to only small corrections, will be effective in performing
quantitative extrapolations to the physical point.  Indeed work in
this direction \cite{Young:2004tb,Leinweber:2004tc,Leinweber:2006ug} has been very
successful.

Comparison of our quenched QCD results with experiment is not as
interesting.  The chiral physics of quenched QCD differs from the
correct chiral physics of full QCD and our results explore
sufficiently light quark masses to reveal these discrepancies.  The
simulation results do not agree with experiment, particularly for
light quark baryons where chiral physics makes significant
contributions.  However, methods have been discovered for
quantitatively estimating the corrections to be encountered in
simulating full QCD and we refer the interested reader to
Refs.~\cite{Young:2004tb,Leinweber:2004tc,Leinweber:2006ug} for further discussion.

In future simulations it will be interesting to explore the utility of
boundary conditions which allow access to arbitrarily small momentum
transfers, providing opportunities to map out hadron form factors in
detail.  Similarly, by calculating near $Q^2 = 0$ one would have more
direct access to the magnetic moment.  Nevertheless, such boundary
conditions cannot be seen to substitute for larger volume lattices, as
the discretisation of the momenta due to the finite volume of the
lattice acts to suppress chiral non-analytic behaviour.  Only with
increasing lattice volumes will the continuous momentum of chiral
loops be approximated well on the lattice.

\acknowledgments

DBL thanks Richard Woloshyn for helpful discussions on the sequential
source technique.  We thank the Australian Partnership for Advanced
Computing (APAC) and the South Australian Partnership for Advanced
Computing (SAPAC) for supercomputer support enabling this project.
This work is supported by the Australian Research Council.

\end{document}